\documentclass[11pt]{article}
\usepackage[margin=1in]{geometry}
\usepackage{subfiles}
\usepackage{standalone}
\usepackage[utf8]{inputenc}
\usepackage{amsthm}
\usepackage{amsmath}
\usepackage{algorithm}
\usepackage{mathtools}
\usepackage{cite}
\usepackage{url}
\usepackage{amsfonts}

\usepackage{bbm}

\usepackage[noend]{algpseudocode}
\algrenewcommand\algorithmicloop{\textbf{repeat}}
\algrenewcommand\algorithmicdo{\textbf{}}
\algrenewcommand\algorithmicend{\textbf{}}

\renewcommand{\algbreak}{\textbf{break }}

\algnewcommand\algorithmicforeach{\textbf{for each}}
\algdef{S}[FOR]{ForEach}[1]{\algorithmicforeach\ #1\ \algorithmicdo}

\usepackage{xcolor}
\newtheorem{theorem}{Theorem}

\newtheorem{lemma}[theorem]{Lemma}

\newtheorem{corollary}[theorem]{Corollary}

\newtheoremstyle{case}{}{}{}{}{}{:}{ }{}
\theoremstyle{case}

\def\Pr{\mathbf{Pr}}

\newcommand{\Prob}[1]{{\Pr\left[\,{#1}\,\right]}}

\newcommand{\defn}[1]{\emph{\textbf{#1}}}
\newcommand{\id}[1]        {\ifmmode\mathit{#1}\else\textit{#1}\fi}

\newcommand{\card}[1]{\left|#1\right|}
\newcommand{\ceil}[1]{\left\lceil#1\right\rceil}

\newcommand{\seclabel}[1]{\label{sec:#1}}
\newcommand{\secref}[1]{Section~\ref{sec:#1}}

\newcommand{\alglabel}[1]  {\label{alg:#1}}
\renewcommand{\algref}[1]    {Algorithm~\ref{alg:#1}}

\newcommand{\lemlabel}[1]{\label{lemma:#1}}
\newcommand{\lemref}[1]{Lemma~\ref{lemma:#1}}
\newcommand{\theoremlabel}[1]{\label{theorem:#1}}
\newcommand{\theoremref}[1]{Theorem~\ref{theorem:#1}}

\title{Efficient Construction of Directed Hopsets and Parallel Approximate Shortest Paths}
\author{
  Nairen Cao\\
  Georgetown University\\
  \texttt{nairen@ir.cs.georgetown.edu}
  \and 
  Jeremy T. Fineman\\
  Georgetown University\\
  \texttt{jfineman@cs.georgetown.edu}
  \ \and 
  Katina Russell\\
  Georgetown University\\
  \texttt{katina.russell@cs.georgetown.edu}
}
\date{}
\begin{document}

\maketitle

\begin{abstract}
  The approximate single-source shortest-path problem is as follows:
  given a graph with nonnegative edge weights and a designated source
  vertex $s$, return estimates of the distances from~$s$ to each other vertex
  such that the estimate falls between the true distance and
  $(1+\epsilon)$ times the distance.  This paper provides the first
  nearly work-efficient parallel algorithm with sublinear span (also called
  depth) for the approximate shortest-path problem on \emph{directed}
  graphs.  Specifically, for constant $\epsilon$ and
  polynomially-bounded edge weights, our algorithm has work
  $\tilde{O}(m)$ and span $n^{1/2+o(1)}$. Several algorithms were
  previously known for the case of \emph{undirected} graphs, but none
  of the techniques seem to translate to the directed setting.

  The main technical contribution is the first nearly linear-work
  algorithm for constructing hopsets on directed graphs. A
  $(\beta,\epsilon)$-hopset is a set of weighted edges (sometimes
  called shortcuts) which, when added to the graph, admit $\beta$-hop
  paths with weight no more than $(1+\epsilon)$ times the true
  shortest-path distances.  There is a simple sequential algorithm that takes as
  input a directed graph and produces a linear-cardinality hopset with
  $\beta=O(\sqrt{n})$, but its running time is quite
  high---specifically $\tilde{O}(m\sqrt{n})$.  Our algorithm is the
  first more efficient algorithm that produces a directed hopset with similar
  characteristics.  Specifically, our sequential
  algorithm runs in $\tilde{O}(m)$ time and constructs a hopset with
  $\tilde{O}(n)$ edges and $\beta = n^{1/2+o(1)}$.  A parallel version
  of the algorithm has work $\tilde{O}(m)$ and span $n^{1/2+o(1)}$.
\end{abstract}
\pagebreak

\section{Introduction}
The single-source shortest-path problem on graphs with nonnegative
edge weights is notoriously difficult to parallelize.\footnote{Perhaps
  counter-intuitively, achieving at least a reasonable level of
  parallelism when the weights are both positive and negative is
  easier. This is in part because algorithms for the general case have
  roughly the same inherent sequential dependencies but with far more
  work that can be parallelized in each step.}  In the sequential
setting, the classic solution has running time
$O(m + n\log n)$~\cite{FredmanTa87}, where throughout $n$ denotes the
number of vertices and $m$ the number of edges.  Given that the
sequential solution has nearly linear runtime, an ideal parallel
algorithm would run in $\tilde{O}(m/p)$ parallel time on $p$
processors (for large~$p$), where the $\tilde{O}$ notation suppresses
logarithmic factors.  Achieving such a bound requires a parallel
algorithm with nearly linear work and strongly sublinear span; the
\defn{work} of a parallel algorithm is the total number of primitive
operations, and its \defn{span} is the length of the longest chain of
sequential dependencies or equivalently the limit of the parallel time
as $p$ approaches infinity.  The exact version of the shortest-path
problem is well-studied (see
e.g.~\cite{DriscollGaSh88,BrodalTrZa98,MeyerSa03,UllmanYa91,Spencer97,Klein:1997:RPA:266956.266959,BlellochGuSu16}),
but no ideal parallel solutions exist, especially when the graph is
sparse.\footnote{Achieving parallelism $p=O(m/n)$ is fairly
  straightforward.}  Even for the simplest case of a unweighted,
undirected graph, all algorithms to date either have linear span,
meaning that they are inherently sequential, or they reduce the span
by increasing the work.  For example, when tuned to achieve span of
$\tilde{O}(\sqrt{n})$, Spencer's algorithm~\cite{Spencer97} has work
$\tilde{O}(m+n^2)$ and Ullman and Yannakakis's
algorithm~\cite{UllmanYa91} has work $\tilde{O}(m\sqrt{n})$.

For undirected graphs at least, there has been more success on the
approximate version of the problem.  In the approximate shortest-path
problem with source vertex $s$, the algorithm must output for all
vertices $v$ an estimate $d_v$ on the shortest path-distance such that
$\id{dist}(s,v) \leq d_v \leq (1+\epsilon) \id{dist}(s,v)$, where
$\id{dist}(s,v)$ is the shortest-path distance from $s$ to $v$.
Several algorithms have been designed for this approximate problem on
undirected graphs, see
e.g.,~\cite{Cohen00,MPVX15,ElkinNe16,ElkinNe19}.  These algorithms
exhibit work-span tradeoffs.  It is possible~\cite{MPVX15} to achieve
$O(n^\alpha)$ span, for arbitrarily small constant $\alpha$, while
still maintaining nearly linear work $\tilde{O}(m)$. It is also
possible to achieve even lower span with higher work~\cite{Cohen00,ElkinNe16,ElkinNe19}.  

A natural question is whether it is possible to achieve nearly linear
work and sublinear span for approximate shortest paths on
\emph{directed} graphs.  This paper answers the question in the
affirmative: we present an algorithm for directed graphs with
$\tilde{O}(m)$ work and span $n^{1/2 + O(1/\log\log n)}$.

\subsubsection*{Hopsets} 
While it is unknown how to efficiently compute shortest paths in
general in parallel, it is known how to find the shortest $\beta$-hop
paths.  Specifically, Klein and Subramanian give an exact
algorithm~\cite{Klein:1997:RPA:266956.266959} for integer weights with
work $\tilde{O}(m\log W)$ and span $\tilde{O}(\beta)$, where $W$ is
the maximum integer weight.  This algorithm can also be trivially
extended to a $(1+\epsilon)$-approximate algorithm on nonnegative real
weights by first normalizing so that the minimum nonzero weight is
$1/\epsilon$ then rounding up each weight to the next integer.  Given
Klein and Subramanian's algorithm, the natural approach is to first
preprocess the graph to produce a new graph whose $\beta$-hop
distances are not too much larger than the actual unbounded distances;
the preprocessing step amounts to finding a good hopset.

A \defn{$(\beta,\epsilon)$ hopset} $H$ is a set of weighted edges
that, when added to the original graph, approximates the shortest-path
distances by paths of at most $\beta$ hops, where $\beta$ is called
the \defn{hopbound}.  Formally, let $G=(V,E)$ be the original graph
and $G' = (V,E \cup H)$ be the graph with the hopset edges included.
$H$ is a $(\beta,\epsilon)$ hopset if and only if (1) for all edges
$(u,v) \in H$, the weight $w(u,v)$ of the edge is no lower than the
shortest-path distance in $G$, i.e., $w(u,v) \geq \id{dist}_G(u,v)$,
and (2) there exists a path $p$ from $u$ to $v$ in $G'$ comprising at
most $\beta$ hops such that $w(p) \leq (1+\epsilon) \id{dist}_G(u,v)$.
(The first constraint implies that $w(p) \geq \id{dist}_G(u,v)$.)
Although hopsets were first formalized by Cohen~\cite{Cohen00}, they
were used implicitly in many of the prior algorithms.  Most algorithms
for constructing hopsets, including the one in this paper, are
randomized and there is some small chance that the weight of some
$\beta$-hop path will be too high.

There are several features characterizing the quality of a hopset: the size
or number of edges in the hopset, the hopbound $\beta$, the
approximation quality $\epsilon$, and the complexity of an algorithm
for constructing the hopset.  When $\epsilon = 0$, the hopset produced
is an \defn{exact hopset}, meaning that the $\beta$-hop distances in
the augmented graph are the true shortest-path distances.

There is a simple folklore sequential algorithm for constructing an
exact hopset with hopbound $\beta=\tilde{O}(\sqrt{n})$ and size
$O(n)$.  The algorithm is as follows.  First sample each vertex with
probability $O(1/\sqrt{n})$.  Next, compute the single-source
shortest-path distances from each sampled vertex to all other sampled
vertices.  For samples $s_i$ and $s_j$, add to hopset $H$ the edge
$(s_i,s_j)$ with weight $w(s_i,s_j) = \id{dist}(s_i,s_j)$.  Since
edges are only added between pairs of sampled vertices, the hopset
trivially contains $O(n)$ edges with high probability.  To analyze the
hopbound, consider a shortest path from $u$ to $v$.  With high
probability, the $\beta$ hops nearest to $u$ and $\beta$ hops nearest
to $v$ each contain at least one sampled vertex, so the rest of the
path can by bypassed using a hopset edge.  Ullman and
Yannakakis~\cite{UllmanYa91} and Klein and
Subramanian~\cite{Klein:1997:RPA:266956.266959} give parallel versions
of this algorithm for the unweighted and integer-weighted cases,
respectively.

The preceding algorithm gives an exact hopset with small size and
reasonable hopbound, and it applies to directed graphs as well.  The
problem is that the construction time is too high: the sequential
running time is $\tilde{O}(m\sqrt{n})$ to compute shortest paths from
$\sqrt{n}$ sources. 

For undirected graphs, when the exactness is relaxed and we are
willing to accept a $(1+\epsilon)$ approximation, there exist
linear-size hopsets with much smaller (subpolynomial) hopbound~\cite{ElkinNe19}.  Moreover, there are more efficient
algorithms~\cite{Cohen00,MPVX15,ElkinNe16,ElkinNe19} for constructing
the hopsets.  The algorithms employ clustering techniques that
strongly exploit the symmetry of distances in undirected graphs. 
         
For directed graphs, a hopbound of $O(\sqrt{n})$ is still the best
known for hopsets of linear size, even for approximate hopsets with
large $\epsilon$ and ignoring construction cost.  In fact, if
$\epsilon \geq nW$ and all edge weights are at least one, then
distances themselves become irrelevant---the problem reduces to the
diameter-reduction or shortcutting problem: add edges to the graph,
without changing the transitive closure, to reduce the unweighted
directed diameter, i.e., the number of hops necessary to get from one
vertex to another.  It is yet unknown whether it is always possible to
achieve diameter better than $O(\sqrt{n})$ when restricted to add at
most $n$ edges.  In fact, there is a lower bound of $\Omega(n^{1/6})$
on the diameter~\cite{HuangPe18}, which implies a separation between
the quality of hopsets on directed and undirected graphs.  Revisiting
construction cost, there was no more efficient algorithm known for any
constant $\epsilon$ before the current paper.

\subsubsection*{Our results}

This paper presents the first efficient algorithm for producing a
hopset on directed graphs with sublinear hopbound.  Specifically, our
algorithm produces a $(\beta=n^{1/2+O(1/\log\log n)},\epsilon)$ hopset
with nearly linear size, which is close to matching the quality of the
hopset produced by the highly inefficient folklore algorithm. For
unweighted graphs
(Sections~\ref{section:algorithm}--\ref{section:analysis}), the hopset
has size $\tilde{O}(n/\epsilon^2)$, and the algorithm runs
in time $\tilde{O}(m/\epsilon^2)$. More generally for weighted graphs
(Section~\ref{section:weighted}), the hopset has size
$\tilde{O}(n\log(nW)/\epsilon)$ and the algorithm runs in time
$\tilde{O}(m\log(nW)/\epsilon^2)$, where $W$ is the ratio between the
maximum edge weight and the minimum strictly positive edge weight.  The
construction is successful with high probability, and failure is one
sided---i.e., the result is always a hopset, but the question is
whether it achieves the $(1+\epsilon)$ approximation.

Our parallel algorithm (Section~\ref{section:parallel}) constructs a
hopset with similar characteristics.  The algorithm has work
$\tilde{O}(m\log^2(nW)/\epsilon^4)$ and span $O(n^{1/2+O(1/\log\log n)}/\epsilon)$.

Using our parallel hopset construction then applying Klein and
Subramanian's algorithm~\cite{Klein:1997:RPA:266956.266959} to the
augmented graph yields the first nearly work-efficient parallel
algorithm for approximate single-source shortest paths on directed
graphs with low span. More precisely, our algorithm has work
$\tilde{O}(m\log(nW)/\epsilon^4)$ and span $O(n^{1/2+O(1/\log\log n)}/\epsilon)$.

\subsection{Overview of the algorithm and analysis}

Our algorithm and analysis builds on recent breakthroughs on the
diameter-reduction problem by Fineman~\cite{Fineman18} later improved
by Jambulapati, Liu, and Sidford~\cite{Stanford19}, henceforth
referred to as the JLS algorithm.  This section summarizes the
previous algorithms and key aspects of the analyses, highlights the
difficulties in extending the algorithms to hopsets, and gives an
overview of our insights.  The bulk of this section focuses on the
sequential versions of the algorithms.

The diameter reduction problem is that of adding edges, or shortcuts,
to a directed graph to reduce its unweighted diameter without changing
the transitive closure.  Fineman's algorithm~\cite{Fineman18} is the
first nearly linear-time sequential algorithm with any nontrivial
diameter reduction.  Specifically, his algorithm runs in
$\tilde{O}(m)$ time and creates $\tilde{O}(n)$ shortcuts that reduce
the diameter of any directed graph to $\tilde{O}(n^{2/3})$, with high
probability.  The JLS algorithm~\cite{Stanford19} achieves a diameter
of $n^{1/2+o(1)}$, also with nearly linear running time.  Both
algorithms also have parallel versions with span matching the diameter
achieved to within logarithmic factors.

Our algorithm for hopsets most closely resembles the JLS algorithm for
diameter reduction.

\subsubsection*{Previous algorithms for diameter reduction}

Both Fineman's algorithm~\cite{Fineman18} and the JLS
algorithm~\cite{Stanford19} operate roughly as follows.  Select a
random set of pivots $x_i$; how the pivots are selected varies across
the two algorithms and is discussed more later.  Next perform a graph
search forwards and backwards from each pivot to identify the vertices
reachable in either direction.  Add shortcut edges between the pivots
and all vertices reached, i.e., if a vertex $u$ is reached in backward
direction from pivot $x_i$, then the edge $(u,x_i)$ is added.  Next
partition the vertices into groups according to the set of pivots that
reach them. For example, a group could consist of those vertices
reached by $x_1$ in the forward direction, $x_3$ in the backward
direction, $x_4$ in both directions, and unreached by all other
pivots.  If a group is reached in both directions by the same pivot
(as with the preceding example and pivot $x_4$), mark the group as
done.  Finally, recurse on the subgraph induced by each group that has not been marked as
done.

The main difference between the algorithms is how pivots are selected.
Fineman's algorithm~\cite{Fineman18} selects a single pivot uniformly
at random.  JLS~\cite{Stanford19} instead samples vertices to select a
set of pivots.  The algorithm is parameterized by a value $k$ that
controls the sampling probability; $k = \Theta({\rm poly}(\log n))$ is
a good choice, so we shall assume as much going forward to simplify
the statement of remaining bounds.  Each vertex is a selected as a
pivot with probability $k^{r+\Theta(1)} / n$, where $r$ is the
recursion depth.  The probability of becoming a pivot thus increases
by a factor of $k$ with each level of recursion, and it is possible to
select many pivots.  Beyond achieving a better diameter, the JLS
algorithm also has the advantage that the recursion depth is trivially
limited to $\log_kn$.  Increasing $k$ impacts the total work as
multiple overlapping searches are performed, which is why $k$ should
not be too large.  We shall not discuss the analysis of the running
time here, but suffice it to say that it is not hard to show that
these sequential algorithms have $\tilde{O}(m)$ running time.

The diameter analysis starts by fixing any long $s$-to-$t$ path $P$ to
analyze. The goal is to argue that with at least constant probability,
the addition of shortcuts introduces a short-enough $s$-to-$t$ path to
the graph.  The algorithm can be repeated to boost the success
probability.

One of the key setup ideas is classifying vertices according to how
they relate to the path~$P$.  We write $v \preceq P$ if it is possible
to get from $v$ to any vertex on $P$ by following directed edges and
$P \preceq v$ if it is possible to get from any vertex on $P$ to $v$
by following directed edges.  A vertex $v$ is an \defn{ancestor} of
$P$ if $v\preceq P$ and $P\not\preceq v$. The vertex is a
\defn{descendant} of $P$ if $v\not\preceq P$ and $P\preceq v$.  It
is a \defn{bridge} if $v \preceq P$ and $P \preceq v$.
The vertex is \defn{unrelated} otherwise.  

As the algorithm executes and partitions the graph, so too does it
partition the path being analyzed.  An execution can be modeled by a
recursion tree where only the \defn{relevant subproblems}, i.e., those
that contain subpaths of $P$, are included.  The leaves of this
relevant subproblem tree occur when at least one of the pivots is a
bridge; if a bridge is selected, then edges are added between all
vertices on the subpath and the bridge in both directions, meaning
that the subpath has been shortened to two hops.  The final path
length from $s$ to $v$ is thus upper bounded by the number of leaves
in the tree of relevant subproblems. 

For the case of a single pivot as in Fineman's
algorithm~\cite{Fineman18}, it is not hard to see that a relevant
subproblem gives rise to at most two recursive subproblems, and the
two subproblems occur only if the pivot is an ancestor or descendant.
For example, if the pivot is an ancestor, the path is partitioned at
the first reachable vertex on the path.  If an unrelated pivot is
selected, there is only one relevant subproblem; informally, this case
can be ignored in the single pivot case as tree nodes with a single
child can be contracted.  More generally, JLS show~\cite{Stanford19}
that if $t$ ancestors/descendants are selected, then the path is
partitioned across at most $t+1$ relevant subproblems.

A key component of the analysis is to show that the total number of
ancestors and descendants is likely to decrease each time an ancestor
or descendant pivot is selected.  It thus becomes less and less likely
to partition the path further and more likely to select a bridge.  For
concreteness, let us first consider a sketch of the intuition for the
single-pivot case.  Fineman~\cite{Fineman18} proves that if a random
ancestor is selected as the pivot, then the total number of ancestors
across both recursive subproblems reduces by a factor of $1/2$ in
expectation.  Similarly for descendants.  We thus need roughly
$(1/3)\lg n$ levels of recursion to reduce the total number of
ancestors to $n^{2/3}$ and another $(1/3)\lg n$ levels to similarly
reduce the number of descendants.  At recursion depth $(2/3)\lg n$,
there are thus at most $2^{(2/3)\lg n} = n^{2/3}$ subproblems and at
most $O(n^{2/3})$ ancestors and descendants.  Even if all of the
remaining ancestors and descendants eventually become pivots, there
can be at most $O(n^{2/3})$ leaves in the recursion tree, which yields
the final path length.

If one could ensure that the algorithm always selects either zero or
$t$ related pivots, then one could easily extend Fineman's analysis to
the multi-pivot case.  In particular JLS prove~\cite{Stanford19} that
with $t$ random ancestor/descendant pivots, the total number of
ancestors and descendants reduces by $c/(t+1)$ in expectation, for
some constant~$c$.  Consider the $r$th level of recursion assuming $t$
related pivots are always selected.  The number of subproblems is at
most $(t+1)^r$.  The number of ancestors and descendants is upper
bounded by $c^rn/(t+1)^r$, which also upper bounds the number of
leaves that could arise lower in the recursion tree.  Setting
$r = (1/2)\log_{t+1} n$ roughly balances these two terms and gives a path
length of at most
$\sqrt{n}c^{\log_{t+1} n} = n^{1/2 + O(1/\log(t+1))}$.

Unfortunately, the algorithm is unaware of the path $P$, and it cannot
ensure that $t$ of the pivots are related to the path. Nevertheless it
is still possible to obtain the same bound. The JLS
analysis~\cite{Stanford19} adopts a bottom-up approach, solving a
recurrence on the shortcutted path length for a given number of
ancestors/descendants.

\paragraph{Parallel versions.} The big challenge in parallelizing
these algorithms is performing the graph searches used to partition
the graph.  To achieve low span both Fineman and JLS employ
$h$-hop-limited searches, i.e., only identifying vertices reachable
from the pivot within $h$ hops.  Fineman and JLS set $h$ to
$h=\tilde{\Theta}(n^{2/3})$ and $h=n^{1/2+o(1)}$, respectively.  As
noted previously, there are parallel algorithms implementing $h$-hop
limited searches with $\tilde{O}(h)$
span~\cite{Klein:1997:RPA:266956.266959}.  Unfortunately, using
hop-limited searches it is no longer immediately true that selecting
$t$ related pivots partitions the path into at most $t+1$ subpaths,
which was crucial for the analyses.  To fix this issue,
Fineman~\cite{Fineman18} and JLS~\cite{Stanford19} (1) only analyzes
paths with length $\tilde{\Theta}(h)$, and (2) handle vertices near
the boundary of the search, called \defn{fringe vertices}, differently
from other vertices.  In doing so, they are able to achieve the
$\leq t+1$ relevant subproblems, though the details become
significantly more complicated.

\subsubsection*{Key challenge for hopsets}

A natural first step to extend the diameter-reduction algorithms to 
build hopsets is to add weights to any added shortcuts.  Specifically,
perform a shortest-path algorithm from each pivot and augment the shortcuts
with weight equal to the shortest-path distances to each vertex.   Our
algorithm includes weights on shortcuts, but this change alone is not sufficient to achieve a good approximation.

The main challenge is that bridges do not necessarily make good
pivots.  Specifically, consider any bridge $x$ for an $s$-to-$t$ path.
If $x$ is selected as a pivot, then a 2-hop path is created from $s$
to $t$, which is enough for the diameter-reduction problem.  For
hopsets, however, the weight of the path matters.  If
$\id{dist}(s,x) + \id{dist}(x,t) \gg \id{dist}(s,t)$, then the 2-hop
path taking the shortcuts does not approximate the shortest-path
distance.  It may thus be necessary to continue recursing on subpaths
in subproblems until better shortcuts are found.

The challenge becomes more prominent when we start to dissect the analysis.
We focus on the single-pivot case of Fineman's
algorithm~\cite{Fineman18} here for simplicity.  Recall the two key
components of the analysis: (a) selecting a random ancestor (or
descendant) reduces the number of ancestors (or descendants) by $1/2$
in expectation, and (b) selecting a random ancestor or descendant
partitions the path across at most two subproblems.  Selecting a
far-away bridge as pivot falls short in both respects.  It is not
possible to argue anything analogous to (a)---it could be that for
every bridge, few ancestors, descendants, or bridges are knocked out.
Moreover, selecting a bridge as pivot may give rise to three, not just
two, relevant subproblems: the path can comprise a subpath reached by
the backward search followed by an unreached subpath followed by a
subpath reached by the forward search.

\subsubsection*{Our algorithm for hopsets}

Our algorithm builds off the JLS algorithm, also parameterized by
sampling parameter $k$, but with several key modifications.  The goal
is to circumvent the preceding challenge by ensuring, at least in
effect, that shortcuts added to or from bridges are good enough
for the approximation.  We first summarize the differences in the
algorithm before revisiting the analysis.

\begin{enumerate}
\item
\textbf{Pivots and shortcutters.} In the previous
algorithms, pivots are used both to partition the graph and to add
shortcuts.  Here, we split the roles; we use some vertices, called
\defn{pivots} to establish the partition of the graph, and other
vertices, called \defn{shortcutters}, to add edges to the hopset.
Pivots are selected analogously to JLS, but we sample a larger set of
shortcutters.  More precisely, if a vertex becomes a pivot at
recursion depth $r$, then it first becomes a shortcutter at recursion
depth $r-f(\epsilon,n)$ for some function $f$.  Larger $f$ improves the
approximation quality but increases the work of the algorithm.

\item\textbf{Weighted shortcuts.} From each shortcutter
$s$, we compute the single-source shortest paths from $s$ to all other
vertices in both the forwards (and backwards) directions.  We then add
the weighted edges $(s,v)$ (and $(v,s)$) with weight $w(s,v) =
\id{dist}(s,v)$ (and $w(v,s) = \id{dist}(v,s)$) to the hopset.  Using
weighted shortcuts is the obvious modification necessary for a
hopset.  
\item\textbf{Decreasing distance-limited searches from pivots.} To
  establish the graph partition, we perform graph searches from each
  pivot as before, but the searches are now limited to a bounded
  distance.  Moreover, the search distances decrease by a factor of
  $\lambda\sqrt{k}$ with each level of recursion, for constant
  $\lambda$.  The initial distance is important---the algorithm only
  well-approximates paths if the initial search distance is similar to
  the shortest-path distance---so we run the algorithm at all relevant
  initial-distance scales.
\end{enumerate}

It is worth noting that the distance-limited searches here are not
analogous in purpose to the hop-limited searches used by the
prior~\cite{Fineman18,Stanford19} parallel algorithms for diameter
reduction.  (Our parallel version also imposes a hop limit.)  Here the
distance-limited searches are important even for the sequential
algorithm in order to obtain a good approximation.  Moreover, the
distances decrease significantly with each level of recursion, whereas
the hop-limited searches use roughly the same number of hops at all
levels.  Nevertheless, some of the technical machinery (e.g., fringe
vertices) is similar.  

Because our sequential algorithm for hopsets uses distance-limited
searches, the details of both the algorithm and analysis are more
complicated than the sequential algorithms for diameter reduction.

\subsubsection*{Key ideas of the analysis}

Our analysis has two main novelties, summarized next.  Note that the
bounds stated here are correct in spirit but imprecise in that that
they omit some lower-order terms in favor of conciseness.  

For the following discussion, it is important to interpret the vertex
classifications (ancestor, descendant, and bridge) to be with respect
to the bounded distances, analogous to the hop-limited searches in
prior work~\cite{Fineman18,Stanford19}.  For example, a vertex is only
a bridge if it can reach the path in both directions by an appropriate
distance-limited search.

The first technical contribution can be viewed as an alternative way
of analyzing the JLS algorithm, but this version makes it easier to
cope with the new features of the hopset algorithm.  Specifically, we
show that the number of subproblems increases by at most $O(\sqrt{k})$
on average with each level of recursion.  For any constant in the
big-O, it follows that there be at most
$(O(\sqrt{k}))^r = (k^{1/+O(1/\log k)})^r$ relevant subproblems at
recursion depth~$r$.  Looking at the maximum recursion depth
$r=\log_k n$ gives a direct bound of $n^{1/2+O(1/\log k)}$ on the
number of relevant subproblems, and hence the length in hops of the
shortcutted path.

Now consider what happens if we augment the JLS algorithm with
decreasing distance-limited searches.  Let $w(P)$ be the weight of the
path $P$ being analyzed, and assume that the initial search distance
is roughly $w(P)$.  The general issue when decreasing the search
distance is that when searches do not reach the end of the path, the
path may be partitioned into more pieces than desired.\footnote{The
  use of ``fringe vertices'' suffices if the search distance is
  sufficiently long with respect to the path length.  The new issue
  that arises here is that the search distance can be significantly
  shorter than the path length.}  We circumvent the issue by logically
dividing any long paths into subpaths of length roughly
$w(P)/(\lambda\sqrt{k}))^{r}$ (proportional to the search distance),
where $r$ is the recursion depth. In this way, the searches can now
traverse the full length of the path.  It is easy to see that there
can be at most $O((\lambda\sqrt{k})^{r})$ logical subproblems created.
For large-enough $\lambda$, this term dominates the number of
subproblems arising from the previous level of recursion, so we have a
total of $O(\lambda^rk^{r/2})$ subproblems at recursion depth~$r$.
Again, this bound readily implies a hop bound for the shortcutted path
of $n^{1/2+O(1/\log k)}$, albeit with a larger constant in the big-O.

The second new idea is in analyzing the approximation factor achieved
by the hop set, which requires all three algorithmic modifications.
Let us first consider only the shortcuts generated by ``nearby''
bridges. For an $s$-to-$t$ subpath at recursion depth $r$, we say that
a bridge $x$ is nearby if
$\id{dist}(s,x)+ \id{dist}(x,t) = \id{dist}(s,t)+O((\epsilon/\log n)
w(P) / (\lambda^rk^{r/2}))$.
Since the total number of subproblems is $O(\lambda^rk^{r/2})$,
shortcuts from nearby bridges contribute a total of
$O(\epsilon w(P)/\log n)$ additive error to the path length.  Summing
across all $O(\log n)$ levels of recursion gives a total additive
error of $O(\epsilon w(P))$, and hence a multiplicative error of
$(1+O(\epsilon))$.

The goal is thus to show that all bridges are effectively nearby
bridges.  This statement seems implausible, but we can achieve it by
leveraging both the bounded search distance as well as the
oversampling of shortcutters.  In fact, for
$\epsilon = \Omega(\log n)$, we can immediately see that all bridges
are nearby---the additive error is bounded by twice the maximum search
distance, i.e.,
$O(w(P)/(\lambda^rk^{r/2})) = O((\epsilon /\log n)
w(P)/(\lambda^rk^{r/2}))$.  We thus achieve a hopset with $\epsilon =
O(\log n)$ even setting the shortcutters and pivots to be identical.

To achieve a better approximation, we leverage the oversampling of
shortcutters.  Observe that moving the shortcutters to a higher level
of recursion can only improve the length in hops of the shortcutted
path, as strictly more edges are added.  To analyze quality of the
approximation, we consider the recursion tree of relevant subproblems,
but we now have a base case whenever a nearby bridge is selected as a
shortcutter.  

Since moving shortcutters higher in the recursion only helps, it
suffices to show that the pivots selected in relevant subproblems are
never bridges, i.e., that all shortcuts important to the hopbound
also have small additive error.  We prove the claim that pivots are
never bridges by contradiction.  Suppose that a pivot $x$ is a bridge
in a relevant subproblem at recursion depth~$r$.  Then it must be
within a distance $O(w(P)/(\lambda^rk^{r/2}))$ of both the start and
end of the path, as that is both the search distance and the path
length.  The additive error contributed by this bridge is thus at most
$O(w(P)/(\lambda^rk^{r/2}))$.  While $x$ would not be considered a
nearby bridge at level $r$, recall that $x$ is first selected as a
shortcutter at recursion depth $r-f(\epsilon,n)$.  For appropriate
choice of $f$, i.e.,
$(\lambda\sqrt{k})^{f(\epsilon,n)} = \Omega(\log n / \epsilon)$, $x$
is a nearby bridge at depth $r-f(\epsilon,n)$, constituting a base
case of the recursion.  Thus the subproblem in which $x$ is selected
as a pivot is not a relevant subproblem.

\section{Preliminaries}
A directed weighted graph is a pair $(G, w)$ where $G = (V, E)$ is a graph and $w: E \rightarrow R$ is a weight function. In this paper, we treat $w$ as an attribute of $E$. Hence, we refer $G$ as the weighted graph and ignore $w$. For a weighted directed graph $G = (V, E)$, the number of vertices and edges are $|V| = n $ and $|E| = m$, respectively. For $e \in E$, we denote the weight as $w_E(e)$ and we write $w_E(e)$ as $w(e)$ for simplicity. If $e \not\in E$, then $w(e) = +\infty$. If the graph is unweighted, then $w(e) = 1$ for all $e \in E$.  For a subset $V' \subset V$, we denote the induced graph on $V'$ as $G[V']$. 
For any vertices $u, v \in V$, define $\id{dist}_G^{(\beta)}(u, v)$ to be the minimum weighted path from $u$ to $v$ containing at most $\beta$ edges. If there is no path containing at most $\beta$ edges from $u$ to $v$, then $\id{dist}_G^{(\beta)}(u,v) = +\infty$.  
We also refer to $\id{dist}_{G}(u, v)$ as the shortest path from $u$ to $v$. 
For a set of edges $E$ and a constant $c$, we define $c \cdot E$ to be $E$ where the weight of each edge in $E$ is multiplied by $c$. 
For two set of edges $E$ and $E'$, the union of $E$ and $E'$ is denoted as $E \cup E' = \{e | e \in E \textbf{ or } e \in E'\}$ and the weight function for $e \in E \cup E'$ is the minimum weight of $w_E(e)$ and $w_{E'}(e)$, i.e, $w_{E \cup E'}(e) = min( w_{E}(e), w_{E'}(e))$. 
We assume the lightest non-zero edge weight is 1, and the heaviest edge weight is $W$.  
If the lightest non-zero edge weight $w(e)$ is less than 1, then all edges are scaled by $1/w(e)$.

\paragraph{Paths.}
A \textbf{path} $P = \langle v_0, v_1, \ldots v_\ell \rangle$ is a sequence of constituent vertices such that $(v_i,v_{i+1})$ is an edge in the graph, for all $i \in [0, \ell - 1]$. We denote the length of path $P$ as $|P|$ and $|P| = \ell$ is the number of edges on $P$. 
We also call $|P|$ the number of hops of $P$. 
The first and the last vertex of the path are $\id{head}(P) = v_0$ and $\id{tail}(P) = v_\ell$. 
For a vertex $v$, we say $v \in P$ if $v = v_i$ for some $i \in [0, \ell]$.
We consider the weight of path $P$ to be the sum of the weights of the edges that make up the path, $w(P) = \sum_{i=1}^\ell w(v_{i-1},v_i)$. 
A path $P'$ is a $(1+\epsilon)$-approximation path for another path $P$, if $\id{head}(P) = \id{head}(P')$, $\id{tail}(P) = \id{tail}(P')$, and $w(P) \leq w(P') \leq (1+\epsilon) w(P)$.

\paragraph{Hopsets.}
A \textbf{$(\beta, \epsilon)-$hopset} for directed graph $G=(V,E)$ is a set of weighted edges $H$, such that for any vertices $u$ and $v$ in $V$, $\id{dist}_{G}(u, v) \leq \id{dist}^{(\beta)}_{G'}(u, v) \leq (1 + \epsilon)\id{dist}_{G}(u, v)$, where $G' = (V, E \cup H)$.  
$\beta$ is considered the $\textbf{hopbound}$ of the hopset, and $\card{H}$ is the size of the hopset.

\paragraph{Related nodes.}
For nodes $u,v$ define the relation $u \preceq_d v$ if and only if $\id{dist}_G(u, v) \leq d$. We say $u$ can reach $v$ within $d$-distance or $v$ can be reached by $u$ within $d$-distance if $u \preceq_d v$. If $u \preceq_d v$ or $u \preceq_d v$, then $u$ and $v$ are \textbf{$d$-related}. For a directed graph $G = (V, E)$ and vertices $u,v \in V$, denote $R^{+}_d(G, v) = \{u | v \preceq_d u\}$ and $R^{-}_d(G, v) = \{u | u \preceq_d v\}$ be the set of nodes which can be reached by $v$  and which $v$ can reach within $d$-distance. We denote the set $R_d(G, v) = R^{+}_d(G, v) \cup R^{-}_d(G, v)$ be $v$'s related nodes within $d$-distance. If $d = n$, we will ignore $d$. Similarly, we can define $R^{+}_d(G, P) = \{u | v_i \preceq_d u, v_i \in P\}$,  $R^{-}_d(G, v) = \{u | u \preceq_d v_i \in P\}$ and $R_d(G, P) = R^{+}_d(G, P) \cup R^{-}_d(G, P)$. If $v \in R_d(G, P)$, then $v$ and $P$ are \textbf{$d$-related}.

\paragraph{Path related nodes.}
 For a vertex $x$ and a path $P$, $x$ is a \textbf{d-descendant} of $P$ if and only if $x \in R^{+}_d(G, P) \backslash R^{-}_d(G, P)$.  Vertex $x$ a \textbf{d-ancestor} of $P$ if and only if $x \in R^{-}_d(G, P) \backslash R^{+}_d(G, P)$. $x$ a \textbf{d-bridge} of $P$ if and only if $x \in R^{-}_d(G, P) \cap R^{+}_d(G, P)$. Notice that these sets are all disjoint by definition.

\paragraph{Binomial distribution.} In the paper, denote binomial variables with $n$ independent experiments and probability $p$ as $B(n, p)$. For a random variable $X$, if $X \sim B(n, p)$, the following holds by a Chernoff bound,
\begin{align*}
    \Prob{X \geq (1 + \delta)np} \leq exp(-\frac{\delta^2}{2 + \delta}np).
\end{align*}
If $X \sim B(n, p)$, then 
\begin{align*}
    E[\frac{1}{X + 1}] \leq \frac{1}{E[X]}.
\end{align*}

\section{Algorithm} \label{section:algorithm}

In this section, we describe the hopset algorithm \textsc{Hopset}$(G)$.  
The algorithm takes as input graph $G= (V,E)$, and has parameters $k, \lambda$ and $L$.  
The goal of the algorithm is to output a set of edges $E'$ that is a $(\tilde{O}(n^{1/2}),\epsilon)$-hopset of $G$.

At a high level, the algorithm chooses vertices, called \textbf{pivots}, to search forwards and backwards from adding labels to each reached vertex.  The labels are used to partition the graph into subgraphs for recursion.
There is another set of vertices, called \textbf{shortcutters} that search forwards and backwards adding edges to the hopset for each reached vertex.  The edges that are added to the hopset are weighted by the distance between the shortcutter and the reached vertex. 
The search is limited in distance, so vertices on the boundary of the search, called \textbf{fringe vertices}, are replicated and put into multiple subproblems.  
With each level of recursion, the number of pivots increases, while the search distance decreases. 
The union of the edges added in each level of recursion is returned as the hopset.
Next, we will describe some components of the algorithm and then describe the details of the algorithm.

\paragraph{Parameters.}
The algorithms \textsc{Hopset} and \textsc{HSRecurse} have parameters $k$, $\lambda$ and $L$.  
The parameter $k$ controls the probability that a vertex is chosen as a pivot in each level of recursion. 
The parameter $L$ controls the number of shortcutters in each level of recursion.  
A higher value for $L$ gives a better approximation but also increases the runtime.
Finally, the parameter $\lambda$, which is a constant and controls the probability the algorithm succeeds. The algorithm requires that $k \geq 2$, $\lambda \geq 8$.

\paragraph{Pivots and shortcutters.}
A vertex $v$ is a \textbf{pivot} at recursive level $r$ if $\ell(v) = r$.  
A vertex $u$ is a \textbf{shortcutter} at recursive level $r$ if $\ell(u) \leq r+L$. 
Since each vertex $v$ is assigned $\ell(v)$ at the onset of the algorithm and not changed, we can note that if $v$ becomes a pivot at level $r$, then it was a shortcutter at level $max(0, r-L)$.  
Pivots add labels that partition the graph to each reached vertex in their search.  
Shortcutters add hopset edges but do not add labels, and therefore do not affect the partitioning of the graph at that level.

\paragraph{Search distances.}  Each level of recursion has a range for search distances.  The ranges are disjoint and decreasing with each level of recursion. 
For a level of recursion $r$ and vertex $v$, the search distance is $\rho_vD_r$ where $D_r = D / (\lambda^rk^{r/2})$ is the basic search distance and $\rho_v$ is the scalar. 
The range of search distances is $(\rho_{min}D_r,  \rho_{max}D_r)$, where $\rho_{min} = 16\lambda^2k^2\log^2n - 1$ and $\rho_{max} = 32\lambda^2k^2\log^2n $. 
The searching distance range is divided into $4\lambda^2k\log^2 n$ disjoint subintervals, each with length $4k$. 
A subinterval is chosen uniformly at random, which is represented by $\sigma_v$ in the algorithm.
 Finally, the scalar $\rho_v$ is chosen from within the subinterval to minimize the number of fringe vertices when using search distance $\rho_v D_r$.
We use these search distances to guarantee that there are not too many fringe vertices.

\paragraph{Explanation of Algorithm \ref{hopset} and Algorithm \ref{hsrecurse}.}
\textsc{Hopset}$(G)$, shown in \algref{hopset}, 
repeats $\log n$ times to make the probability of success high.
It assigns $\ell(v) = i$ for each vertex $v$ with a probability $(\lambda k^{i+1} \log n) /n$.  
The $\ell(v)$ is the level of recursion that $v$ becomes a pivot.  
The probability increases by $k$ with each level of recursion.
The recursive subroutine \textsc{HSRecurse}$(G,D,r)$ is called for $D$ set to powers of 2 from $1$ to $\log n$.  
This ensures that a path of any length in $n^{1/2}$ to $n$ is shortcutted.
For each vertex $v$, after assigning $\ell(v)$, if $\ell(v) \leq L$, search forwards and backwards for $2^{j+1}$ and add an edge to the hopset for each reached vertex with weight equal to the distance from the shortcutter to the reached vertex.  Recurse on the whole graph $G$ with $D$ set to $2^jk^{-c}$ for $j \in [1,\log n]$.  Return the set of edges added to the hopset in all recursive executions.

\textsc{HSRecurse}$(G,D,r)$ is the recursive subroutine shown in \algref{hsrecurse}.  It takes as input, a graph $G$, distance $D$, and level of recursion $r$.
For each pivot at level $r$, i.e. each vertex $v$ where $\ell(v) = r$, choose a $\sigma_v$ uniformly at random from $[1,4\lambda k \log^2 n]$. 
Next, search from $v$ to $16\lambda^2k^2 \log n + 4 k \sigma_v$ and find the distance $\rho_v$ that has the minimal number of vertices exactly $\rho_v$ distance away, where $\rho_v$ is limited to restricted to $[16 \lambda^2 k^2\log^2 n + 4k(\sigma_v - 1), 16 \lambda^2 k^2 \log^2 n + 4k\sigma_v)$. Search forwards and backwards from $v$ to distance $\rho_v D_r$ and add labels $v^{Des}$ and $v^{Anc}$ to the vertices reached in the forwards and backwards directions, respectively.  Add the label $X$ on any vertex that is reached in both directions.
Next define the fringe vertices $V_v^{fringe}$ for vertex $v$ as $R_{(\rho_v +1)D_r}(G,v) \backslash R_{(\rho_v -1)D_r}(G,v)$, and recurse on the induced subgraph $G[V_v^{fringe}]$.

Next for each shortcutter, i.e. each vertex $v$ where $\ell(v) \leq r+L$,   search forwards and backwards from $v$ for $32\lambda^2k^2D_r\log^2 n$ and for each reached vertex $u$, add edge $(u,v)$ for ancestors (or $(v,u)$ for descendants) with weight $\id{dist}(u,v)$ (or $\id{dist}(v,u)$ to the hopset.
Next, remove any vertices that received a label $X$ from the pivots.  Finally, partition the vertices into groups as described in the next section, and recurse on the subgraph induced on each group of vertices.

\paragraph{Partition based on labels.}
Line \ref{partitionline} from \algref{hsrecurse} is as follows.  
Partition the graph such that two vertices $u$ and $v$ are in the same group $V_i$, if and only if $u$ and $v$ receive the all the same labels from all pivots.  There could be a group of vertices that receives no labels from any pivots.
Notice that any vertices that received a $X$ label from a pivot are removed in the step before.  
Therefore, none of the subgraphs contain vertices that received a $X$ label.  Finally, the pivots themselves are removed from the graph, as each pivot receives the $X$ label from itself.

\begin{algorithm*}
\caption{Hopset algorithm for unweighted directed graphs. $k, \lambda$ and $L$ are a parameters.}
\alglabel{hopset}
\label{hopset}
\begin{algorithmic}[1]
\Function {Hopset}{$G=(V,E)$}
    \State $H \leftarrow \emptyset$
    \Loop{ $\lambda \log n$ times}
    \ForEach{$j \in [\log n /2,\log n]$}
        \ForEach {$v \in V$}
            \ForEach{$i \in [0, \log_kn]$}
                \State With probability $(\lambda k^{i+1}\log n)/n$, set $\ell(v)$ to $i$, \algbreak if setting successful.
            \If{$\ell(v) \leq L$}
                \ForEach{$u \in R^{+}_{2^{j+1}}(G, v)$} add edge $(v, u)$ to $H$ with weight $\id{dist}_G(v, u)$ 
                \EndFor
                \ForEach{$u \in R^{-}_{2^{j+1}}(G, v)$} add edge $(u, v)$ to $H$ with weight $\id{dist}_G(u, v)$ 
                \EndFor
            \EndIf
            \EndFor
        \EndFor
        \State $H \leftarrow H \cup \Call{HSRecurse}{G, D=2^jk^{-c},r = 0}$
    \EndFor
    \EndLoop
    \State \Return{$H$}
\EndFunction
\end{algorithmic}
\end{algorithm*}

\begin{algorithm*}
\caption{Recursive subroutine for \textsc{Hopset} Algorithm.  $k$, $\lambda$ and $L$ are parameters.}
\alglabel{hsrecurse}
\label{hsrecurse}
\begin{algorithmic}[1]
\Function{HSRecurse}{$G, D, r$}
    \State $D_r \leftarrow D/(\lambda^rk^{r/2}), H \leftarrow \emptyset$
    \ForEach{$v \in V$ with $\ell(v) = r$}
        \State Choose $\sigma_v$ uniformly at random from $[1,4\lambda^2 k\log^2 n]$
        \State Minimize $|R_{(\rho_v+1)D_r}(G,v) \backslash R_{(\rho_v-1)D_r}(G,v)|$  such that $\rho_v \in [16 \lambda^2 k^2\log^2 n + 4k(\sigma_v - 1), 16 \lambda^2 k^2 \log^2 n + 4k\sigma_v)$ 
        \ForEach{$u \in R^+_{\rho_v D_r}(G, v)$} add label $v^{Des}$ to vertex $u$
        \EndFor
        \ForEach{$u \in R^-_{\rho_v D_r}(G, v)$} add label $v^{Anc}$ to vertex $u$
        \EndFor
        \ForEach{$u \in R^+_{\rho_v D_r}(G, v) \cap R^-_{\rho_v D_r}(G, v)$} add label $X$ to vertex $u$
        \EndFor
        \State $V^{\textbf{fringe}}_v \leftarrow R_{(\rho_v +1)D_r}(G,v) \backslash R_{(\rho_v -1)D_r}(G,v)$
        \State $H \leftarrow H \cup \Call{HSRecurse}{G[V^{\textbf{fringe}}_v], D, r+1}$
        \EndFor
    \ForEach{$v \in V$ with $\ell(v) = r+L$}
        
        \ForEach{$u \in R^{+}_{32\lambda^2k^2D_r\log^2 n}(G, v)$} add edge $(v, u)$ to $H$ with weight $\id{dist}_G(v, u)$ 
        \EndFor
        \ForEach{$u \in R^{-}_{32\lambda^2k^2D_r\log^2 n}(G, v)$} add edge $(u, v)$ to $H$ with weight $\id{dist}_G(u, v)$ 
        \EndFor

    \EndFor
    \ForEach{$u \in V$ that has a $X$ label,} {remove $u$} \EndFor
    \State $V_1, V_2,..., V_t \leftarrow$ partition based on labels \label{partitionline}
    \ForEach{$i \in [1, t]$}
        \State $H \leftarrow H \cup \Call{HSRecurse}{G[V_i], D , r+1}$
    \EndFor
    
    \State \Return{$H$}
\EndFunction
\end{algorithmic}
\end{algorithm*}

\section{Analysis} \label{section:analysis}
The goal of this section is to prove the following theorem.
\begin{theorem}
\label{unweightedmaintheorem}
There exists a randomized sequential algorithm that takes a directed graph $G=(V,E)$ where $n=\card{V}$ and $m = \card{E}$, computes a $( n^{1/2 + o(1)}, \epsilon)$-hopset of size $\tilde{O}(n/\epsilon^2)$ with high probability, and runs in $\tilde{O}(m/\epsilon^2)$ time.
\end{theorem}

\noindent 
We start by proving the runtime and the size of the hopset in Section \ref{section:runtime}.  Then we show the hopbound in Section \ref{section:hopbound}, and finally, the approximation in Section \ref{section:approximation}.

\subsection{Running Time and Hopset Size} \label{section:runtime}

In this section we bound the runtime of the algorithm and the size of the hopset the algorithm returns. 
\begin{theorem}\label{runtimetheorem}
One execution of $\textsc{Hopset}(G=(V,E))$ with parameter $k$, where $n=|V|$, $m=|E|$, runs in $\tilde{O}(mk^{L+1})$ time and produces a hopset of size $\tilde{O}(nk^{L+1})$.
\end{theorem}

The proof of Theorem \ref{runtimetheorem} follows the same structure as the runtime proof from JLS \cite{Stanford19}.  First, we bound the related vertices in each recursive subproblem in \lemref{relatedvertices}.  
Then we show the number of times a vertex is added to the fringe problem is small in \lemref{fringevertices}.  
Since only fringe vertices are duplicated, we can bound the total number of vertices and edges in all recursive subproblems in \lemref{vertices-executions}.  This allows us to prove the number of edges added to the hopset and the cost of all recursive executions.  
The runtime differs from JLS \cite{Stanford19} because of the extra searches from shortcutters.  For the same reason, the size of the hopset is larger than the number of shortcutters added in JLS \cite{Stanford19}. 

We start by bounding the number of related vertices in recursive subproblems.  In each level of recursion, the probability of being a pivot increases.  With more pivots, the graph is partitioned into more subproblems, and the number of related vertices in each subproblem decreases.  The proof of vertices in core problems is the same as JLS \cite{Stanford19}.  Our algorithm differs from JLS \cite{Stanford19} for the fringe problem because we increase $r$ as we recurse on fringe problems.  Since the search distance is chosen to minimize the number of vertices on the fringe, the number of vertices in the fringe problem is small, and therefore each vertex does not have too many related nodes.  The upper bound for the vertices in the fringe problem is needed for the hopbound in \secref{hopbound}.

\begin{lemma}\lemlabel{relatedvertices}
Consider an execution of $\textsc{HSRecurse}(G', d, 0)$ on $n$-node $m$-edge graph $G$. With probability at least $1-n^{-0.7\lambda+3}$ in each recursive call of $\textsc{HSRecurse}(G', D, r)$ the following holds for all $v \in G'$,
\begin{equation*} 
|R^{+}_{\rho_{max}D_r}(G', v)| \leq n k^{-r}, |R^{-}_{\rho_{max}D_r}(G', v)| \leq n k^{-r}.
\end{equation*}
\end{lemma}

\begin{proof} 
Proof by induction on $r$.
We will show $R^{-}_{\rho_{max}D_r}(G', v) \leq nk^{-r}$, and $R^{+}_{\rho_{max}D_r}(G', v) \leq nk^{-r}$ follows by a symmetric argument.
For $r=0$, it is clear that the number of related ancestors is at most $n$.

For $r > 0$, there are core problems called by \textsc{HSRecurse}$(G[V_i],D,r+1)$ and fringe problems called by \textsc{HSRecurse}$(G[V_v^{fringe}], D, r+1)$. 
We will start with the related ancestors in the fringe problems, and then show the core problems case.

Each fringe problem is called by some \textsc{HSRecurse}$(G[V_v^{fringe}],D,r+1)$.  Let vertex $u$ be the pivot that calls this fringe problem.  By the inductive hypothesis, $R_{\rho_{max}D_r}(G',u) \leq 2nk^{-r}$.  
The pivot $u$ chooses $\rho$ in 
$[16\lambda^2 k^2\log^2 n + 4k(\sigma_u - 1), 16\lambda^2 k^2\log^2 n + 4k\sigma_u)$, such that the set of vertices $\card{R_{\rho + 1 D_r}(G,v)\backslash R_{(\rho -1)D_r}(G,v)}$ is minimized.  
This size of the range is $4k$, and the search distance for that fringe problem is at most $\rho_{max}$.  By taking the minimum in the range of $4k$, we can guarantee that the total size of the subproblem is reduced by $2/(4k) $, where the 2 comes from $(\rho - 1)D_r$ to $(\rho + 1)D_r$.
Therefore, the size of the problem is at most $nk^{-r-1}$, and each vertex has at most $nk^{-r-1}$ related ancestor or descendant nodes.

Now we will prove the claim for the core problems.  
Let $A =R_{\rho_{min}D_r}^-(G', v)$ be the set of ancestors of $v$ directly after a call of \textsc{HSRecurse}$(G[V_i], D, r + 1)$.  
If $\card{A} > nk^{-r-1}$, then we are done because $\rho_{max}D_{r+1} < \rho_{min}D_{r}$. Otherwise assume $\card{A} \geq nk^{-r-1}$. 
If there are no cycles in $G'$, then the proof is straightforward.  
Order $A$ such that for any $u,y \in A$, if $u \preceq y$ then $u$ precedes $y$.  Call the vertices of $A$ in this order $\langle w_1,w_2,\ldots w_{\card{A}}\rangle$. 
If $w_i$ is a pivot, then any vertex $w_j$, with $j\leq i$ is not in the same subproblem as $v$.
This is because any $w_j$ gets either $w_i^{Anc}$ or no label from $w_i$, whereas $v$ gets $w_i^{Des}$ label.
Therefore, if some $w_i$ is a pivot where $i \geq \card{A} - nk^{-r-1}$, then the number of ancestors of $v$ is no greater than $nk^{-r-1}$.  In level $r$, the probability of being a pivot is $\lambda k^{r+1}\log n / n$.  The probability that no $w_i$ with $i \geq \card{A} - nk^{-r-1}$ is a pivot is
\begin{equation*}
    (1-\frac{\lambda k^{r+1}\log n}{n})^{nk^{-r-1}} \leq e^{-\lambda \log n} \leq n^{-1.4\lambda}.
\end{equation*}

If $G'$ contains cycles, then the proof becomes more complicated. Consider a topologically sorted order of the strongly connected components on $A = \langle A_1, A_2, ... A_i, ... \rangle$.
The difficulty arises within $A_i$ because there is no order between strongly connected vertices.  
It is possible for any order of $A_i = \langle u_1, u_2,... \rangle$, there exists $j$ such that if $u_j$ is a pivot at level $r$, $u_{j'}$ gets $u_j^{Des}$ label and $j' < j$.  
In this case, the argument from the acyclic case no longer holds.  We would like to identify how many nodes in $A_i$ could be in $R_{\rho_{min}D_r}^-(G', v)$. 

We will show later that for any strongly connected components $A_i$, there will be at most $n^{-k-1}/2$ in $R^{-}_{\rho_{max}D_r}(G', v)$ with probability $n^{-.07\lambda + 1}$. 
Now consider the index $j$, such that $|\bigcup_{i > j} A_i| \leq nk^{-r-1}/2$ and $|\bigcup_{i \geq j} A_i| > nk^{-r-1}/2$. 
If $|\bigcup_{i \geq j} A_j| \leq nk^{-r-1}$, then $R_{\rho_{min}D_r}^-(G', v) \leq nk^{-r-1}$ with probability $1 - n^{-0.7\lambda}$. This is the same as acyclic argument except using $n^{-k-1}/2$ instead of $n^{-k-1}$.
Otherwise $|\bigcup_{i \geq j} A_j| > nk^{-r-1}$, in which case $|A_j| \geq n^{-k -1}/2$. 
Based on the strongly connected components argument, $A_j$ will have at most $n^{-k -1}/2$ nodes in $R_{\rho_{min}D_r}^-(G', v)$ and all $A_{j'}$, where $j' < j$, will be not in $R_{\rho_{min}D_r}^-(G', v)$. 
Therefore, $R_{\rho_{min}D_r}^-(G', v) \leq nk^{-r-1}$ with probability $1 - 2n^{-0.7\lambda+1}$.
By taking a union bound over all $v \in V$ and all $r$, the lemma holds for the core problems with probability $1-n^{-0.7\lambda + 3}$.

It remains to prove that for any strongly connected components $A_i$, there will be at most $n^{-k-1}/2$ in $R^{-}_{\rho_{max}D_r}(G', v)$ with probability $n^{-0.7\lambda + 1}$.
For any strongly connected component $A_i$ and two nodes $u_j, u_{j'} \in A_i$, define the relation $\mathcal{R}$ as follows, if $\id{dist}_{G'}(u_j, u_{j'}) \geq \id{dist}_{G'}(u_{j'}, u_{j})$, then $\mathcal{R}(u_j, u_{j'}) = 1$, 
if $\id{dist}_{G'}(u_j, u_{j'}) < \id{dist}_{G'}(u_{j'}, u_{j})$, then $\mathcal{R}(u_j, u_{j'}) = 0$.
  Note that $\mathcal{R}(u_j, u_{j'}) = 0$ implies $\mathcal{R}(u_{j'}, u_j) = 1$.
If $\mathcal{R}(u_{j'}, u_{j}) = 1$ and $u_{j'}$ is chosen as a pivot, then $u_{j}$ gets $u_{j'}^{Des}$ label only if $u_{j}$ gets $u_{j'}^{Anc}$ label. We know that if $u_{j'}$ is chosen as a pivot, $v$ gets only $u_{j'}^{Des}$ label.
Therefore, if $\mathcal{R}(u_{j'}, u_{j}) = 1$ and $u_{j'}$ is chosen as a pivot, then $u_{j}$ is not in $v$'s subproblem.  
Consider the set for node $u_{j} \in A_i$, $T(u_{j}) = \{u_{j'} \in A_i \mid \mathcal{R}(u_{j'}, u_{j}) = 1 \}$.
In other words, $T(u_{j})$ is the set of nodes where if $u_{j'} \in T(u_{j})$ is a pivot means then $u_{j}$ is not in the same subproblem as $v$.
If $|T(u_{j})| > nk^{-r-1}/2$, then $u_{j}$ will be in $R_{\rho_{min}D_r}^-(G', v)$ with probability,
\begin{equation*}
    (1-\frac{\lambda k^{r+1}\log n}{n})^{nk^{-r-1}/2} \leq n^{-0.7\lambda}.
\end{equation*}
Define a set $S = \{ u_j \mid |T(u_j)| \leq nk^{-r-1}/2\}$. 
Based on the above analysis, for a node $u \in A_i$, if $u \not\in S$, then with probability $1 - n^{-0.7\lambda}$, $u \notin R_{\rho_{min}D_r}^-(G', v)$. 
If $u_j \in S$ is a pivot, then there will be at most $nk^{-r - 1}/2$ nodes $u_{j'}$ such that $\mathcal{R}(u_{j'}, u_j) = 1$, which implies there are at most $nk^{-r - 1}/2$ nodes where $\mathcal{R}(u_j, u_{j'}) = 0$.
Therefore, if $u_j$ is a pivot, there will be at most $nk^{-r - 1}/2$ nodes getting $u_j^{Des}$ label and all other nodes in $A_i$ will get $X$ or no label from $u_j$. 

Now we divide the size of $S$ into two cases. If $|S| \leq nk^{-r-1}/2$, then at most $nk^{-r-1}/2$ vertices in $A_i$ will be in $R_{\rho_{min}D_r}^-(G', v)$. On the other hand, if $|S| \ge nk^{-r-1}/2$, a node in $S$ will be a pivot with probability 
\begin{equation*}
    1 -  (1-\frac{\lambda k^{r+1}\log n}{n})^{nk^{-r-1}/2} \geq 1 - n^{-0.7\lambda}.
\end{equation*}
Once there is a pivot in $S$, there will be at most $nk^{-r - 1}/2$ nodes in $R_{\rho_{min}D_r}^-(G', v)$. In both cases, there will be at most $nk^{-r - 1}/2$ nodes of $A_i$ in $R_{\rho_{min}D_r}^-(G', v)$ with probability $1 - n^{-0.7\lambda+1}$. 
\end{proof}

Next, we consider the expected number of nodes added to fringe problems.  JLS \cite{Stanford19} has a similar lemma, where they consider the expected number of times a vertex is added to a fringe problem.  Since we choose a search distance to minimize fringe vertices, we cannot get the same expectation.  Instead, we count the number of vertices each pivot adds to its fringe problem, and get the same result.

The basic search distance $D_r$ for a pivot $v$ is scaled by a factor $\rho_v$ that is chosen from an interval to minimize the number of vertices in the fringe problem.  
The interval that $\rho_v$ is chosen from is selected uniformly at random from a larger interval.  
By using a scaling factor that minimizes the number of vertices on the fringe, and chosen from a random interval, we can guarantee that the number of vertices added to each fringe problem is small.

\begin{lemma}\lemlabel{fringevertices}
Consider a call to \textsc{HSRecurse}$(G',D,r)$ and any vertex $v \in G'$. The expected number of nodes added to $v$'s fringe problem i.e. $\card{R_{(\rho +1)D_r}(G',v) \backslash R_{(\rho -1)D_r}(G',v)}$
is $1/(4\lambda k \log n)$. 
\end{lemma}

\begin{proof} 
If $v$ is not a pivot, we define $v$'s fringe problem size as 0. 
By \lemref{relatedvertices}, the number of related nodes to $v$ is $2nk^{-r}$.
The scaling factor for the search distance,  $\rho$, is chosen from an interval of size $8k$ to select the distance with a minimal number of nodes that are between $(\rho - 1) D_r$ and $(\rho +1)D_r$  distance away.  The $4k$ size interval is chosen uniformly at random from a larger interval of size $4\lambda^2k \log^2 n$.  For a pivot $v$, there will be at most $2nk^{-r} \cdot 1/(4\lambda^2 k \log^2 n) \cdot 2/(4k)$ nodes in its fringe problem in expectation. Multiplying the probability of $v$ being a pivot at level $r$, the expected number of vertices in $v$'s fringe problem is $1/(4\lambda k \log n)$.
\end{proof}

Now that the number of vertices in fringe problems is bounded, we can bound the total number of vertices in all recursive subproblems.  \lemref{vertices-executions} is based on Lemma 5.3 and Corollary 5.5 from JLS \cite{Stanford19}.  The vertices in the core problem form a partition of the vertices in the level before.  The vertices in the fringe problem are copies of vertices in the core problem, which means the total number of vertices increases with each level.  However, since we just showed the number of vertices in the fringe problem is small, the total number of vertices in all recursive subproblems can still be bounded.

\begin{lemma}\lemlabel{vertices-executions}
Consider one execution of \textsc{Hopset}$(G=(V,E))$ where $n =\card{V}$ and $m = \card{E}$.  The expected number of vertices in all recursive executions of \textsc{HSRecurse}$(G',D,r)$ is $2n\log n$.  The expected number of edges in all recursive executions of \textsc{HSRecurse}$(G',D,r)$ is $2m\log n$.
\end{lemma}

\begin{proof}
In one execution of \textsc{HSRecurse}$(G'=(V',E'),D,r)$, the number of vertices called in recursive subproblems is the number called in the fringe problem, \textsc{HSRecurse}$(G[V_u^{fringe}], D, r+1)$, and the number of vertices called in \textsc{HSRecurse}$(G[V_i], D, r+1)$  for $i\in [1,t]$.  By \lemref{fringevertices}, the expected number of nodes added to one vertex's fringe problem is $1/(4\lambda k\log n)$.  
The vertices in $V_i$ are a partition of the vertices in $G'$.  
Therefore the total expected number of vertices in the following subproblem is $\card{V'}(1+1/(4\lambda k\log n))$.  
The total number of levels of recursion is at most $\log_k n$.  Therefore over all levels of recursion, the expected number of vertices in all subproblems is
\begin{equation*}
    \sum_{r=0}^{1+ \log_k n} n (1 + \frac{1}{4\lambda k \log n})^r \leq 2n\log n
\end{equation*}
for $k \geq \log n$.  The edge case can be proved in the same way.
\end{proof}

Next, we bound the number of related pivots each vertex has.  This will set up for the proof of the runtime and size of the hopset.

\begin{lemma}\lemlabel{relatedpivots}
Consider a call to \textsc{Hopset}$(G)$ and all recursive calls of \textsc{HSRecurse}$(G',D,r)$.  For all $v \in V$, the number of pivots $u$, such that $v \in R(G', (\rho_u + 1) D_r, u)$ is $6\lambda k\log n$ with probability at least $1-n^{-0.7\lambda + 4}$.
\end{lemma}

\begin{proof}
To count the number of pivots $u$, where $v \in  R(G', (\rho_u + 1) D_r, u)$, we will slightly overcount the pivots, by extending $\rho_u + 1$ to $\rho_{max}$. This will only increase the pivots we are counting.  
Observe that all pivots $u$ such that $v \in R(G', \rho_{max} D_r, u)$ are in $R(G', \rho_{max} D_r, v)$. By \lemref{relatedvertices}, $\card{R(G', \rho_{max} D_r, v)} \leq 2nk^{-r}$ with probability $1 - n^{-0.7\lambda + 3}$. The number of pivots is a binomial distribution of $B(\card{R(G', \rho_{max} D_r, v)}, \frac{\lambda k^{r+1}\log n}{n})$ and therefore,
\begin{equation*}
    \Pr[B(\card{R(G', \rho_{max} D_r, v)}, \frac{\lambda k^{r+1}\log n}{n}) > 6\lambda k\log n] \leq e^{- 2 \lambda k\log n} \leq n^{-2\lambda}.
\end{equation*}
By taking a union bound over all $v \in V$ and all $r$, the claim holds with probability at least $1 - n^{-0.7\lambda + 4}$.
\end{proof}

\begin{lemma} \lemlabel{relatedshortcutters}
Consider a call to \textsc{Hopset}$(G=(V,E)$. For all $v \in V$, the number of shortcutters $u$, such that $v \in R(G, 2^{j+1}, u)$ is $6\lambda k^{L+1}\log n$ with probability at least $1-n^{-0.7\lambda + 4}$. Consider all recursive calls of \textsc{HSRecurse}$(G',D,r)$.
For all $v \in V$, the number of shortcutters $u$, such that $v \in R(G', \rho_{max}D_r, u)$ is $6\lambda k^{L+1}\log n$ with probability at least $1-n^{-0.7\lambda + 4}$.
\end{lemma}

\begin{proof}
We will prove each of the two cases separately, starting with the second case of the shortcutters in \textsc{HSRecurse}$(G',D,r)$. This case is almost the same as \lemref{relatedpivots} except for the probability of being a shortcutter at level $r$ is $\lambda k^{L+r+1}\log n / n$. Therefore, the expected number of shortcutters $u$ such that $u \in R(G', \rho_{max}D_r, v)$ is $2\lambda k^{L+1}\log n$, and with probability $1 - n^{-0.7\lambda + 4}$, the number of shortcutters $u$ is at most $6\lambda k^{L+1}\log n$.

For the first case, only vertices $v$ where $\ell(v) \leq L$ are shortcutters, and there are at most $n$ vertices. Hence, one vertex is a shortcutter with probability at most $\sum_{i = 0}^{L}\lambda k^{i + 1} \log n /n \leq 2\lambda k^{L+1}\log n /n$, for $k \geq 2$. The number of shortcutters in \textsc{Hopset}$(G)$ is a binomial distribution $B(n, \frac{2\lambda k^{L+1}}{n})$ and by a Chernoff bound, 
\begin{equation*}
    \Pr[B(n, \frac{2\lambda k^{L+1}\log n}{n}) > 6\lambda k^{L+1}\log n] \leq e^{-2\lambda k\log n} \leq n^{-2\lambda}.
\end{equation*}
\end{proof}

Now we can prove Theorem \ref{runtimetheorem}, the runtime of the algorithm, and the size of the hopset.
The runtime is different from the JLS algorithm because of the additional shortcutters that perform searches.

\begin{proof}[Proof of Theorem \ref{runtimetheorem}.]
Assigning probabilities to vertices can be done in linear time.  The searches from pivots and shortcutters can be implemented using a breadth-first search.
The cost of the searches by pivots is the number of edges explored in the breadth-first searches times the number of edges in all recursive subproblems.  
This is $O(mk\log^2 n)$ by \lemref{vertices-executions} and \lemref{relatedpivots}.
Similarly, the cost of searches for shortcutters is the edges explored in the breadth-first searches, which is $6\lambda k^{L+1}\log n$ by \lemref{relatedshortcutters}, times the number of edges in all recursive subproblems, which is $2m\log n$ by \lemref{vertices-executions}.
Finally, the partition step can be implemented to run in $O(n\log nk)$ by sorting different labels.  In total the runtime is $O(mk^{L+1} \log ^4 n)$.
The number of hopset edges added is, at most, the number of vertices explored in the searches.
The total number of vertices searched is the expected number of vertices in all recursive subproblems times the number of times each vertex is searched over all levels of recursion.  
By \lemref{relatedshortcutters} and \lemref{vertices-executions}, this is $O(nk^{L+1}\log^4 n)$ total hopset edges.
\end{proof}

\subsection{Hopbound} \seclabel{hopbound} \label{section:hopbound}
Our goal in this section is to show the hopbound of the hopset produced by the \textsc{Hopset}$(G)$ algorithm is $n ^{\frac{1}{2} + O(1/\log k)}k^{c + \frac{1-L}{2}} \log^2 n$.
%
The main idea comes from Fineman \cite{Fineman18} and JLS \cite{Stanford19}.  
We consider the shortest path $P$ from $u$ to $v$ through the full execution of the algorithm.  If a bridge is selected as a pivot, then the path is shortcutted to two hops.  
If no bridges are selected as pivots, then the pivots are ancestors, descendants, or unrelated the path.  
When an ancestor or a descendant is a pivot, it splits that path into subpaths that are contained in different recursive subproblems. 
Define a \textbf{path-relevant subproblem} $(G, P, r)$ as a call to $\textsc{HSRecurse}(G, D, r)$ that contains a nonempty subpath of $P$.
Splitting the path makes it more challenging to shortcut because a bridge is needed for each subpath in its path-relevant subproblem.  
However, we are still making progress because the number of vertices in path-relevant subproblems is reduced.  
Hence, we would like to track the collection of path-relevant subproblems throughout the execution of the algorithm.  

The path-relevant subproblems form a \textbf{path-relevant subproblems tree}  defined as follows.  
The root of the tree, called level 0, is the whole path $P$.  
If a bridge is selected as a pivot in a path-relevant subproblem, then the node is a leaf and has no children. 
If no bridges are selected in a path-relevant subproblem $(G',P',r)$, then the path-relevant subproblems containing subpaths of $P'$ are the children.
At the end of the execution of the algorithm, the leaves of path-relevant subproblems tree represent the entire path $P$.  
The path consists of at most two hops for each leaf node in the tree and the edges that go between subproblems.  Our goal is to bound the number of nodes in the path-relevant tree to provide an upper bound of the hopbound.  The idea of the path-relevant subproblems tree comes from Fineman \cite{Fineman18}.  However, ours becomes more complicated because we use multiple pivots and fringe and core problems.

In \lemref{core}, we will construct the path-relevant subproblems tree.  The proof relies on a helper lemma to show that choosing ancestor and descendant pivots will decrease the number of path-related nodes.  We will show this claim after \lemref{core} in \lemref{relateddecreases}.  The construction of the path-relevant subproblems tree becomes more complicated for two reasons.  
First, the basic search distance $D_r$ decreases with each level of recursion, which means that a pivot may not reach the end of the path in its search.  
This splits the path into an additional subpath.  
Second, the algorithm calls core and fringe problems from each pivot.  It creates many subproblems, so we must choose which of these subproblems to consider in the analysis.

To resolve the first difficulty, we will logically split certain path-relevant subproblems to create \textbf{logical path-relevant subproblems}.
The path is split logically for the sake of analysis. However, the algorithm is unaware of these splits.  
This means that some logical subproblems are in the same call of \textsc{HSRecurse}$(G, D, r)$, but this will not change our analysis.
Notice that between two consecutive levels, the basic search distance will decrease by a $O(\sqrt{k})$ factor.  The pieces of the subpath are split such that the length of each piece is less than the next level's search distance.  This guarantees that the search distance in the next level is long enough to reach the end of the subpath in the logical path-relevant subproblem.
The ancestors and descendants of each piece of the subpath are copied and added to each relevant subproblem.
By splitting subproblems, we introduce an additional $O(\sqrt{k})$ subproblems, as well as multiple copies of many vertices. 
Fortunately, since we have already shown that path related nodes in one subproblem are bounded, this increase in vertices is tolerable.

More specifically, each call to $\textsc{HSRecurse}(G, D, r)$ is associated with path $\hat{P}$ where $|\hat{P}| = \ell \in ( k^c D / 2, k^c D]$.
If a path-relevant subproblem $(G', P, r)$ at level $r$ contains a subpath $P = \langle v_{i}, v_{i+1}, ..., v_{j}\rangle$ with $j - i > D_r = \frac{D}{\lambda^r k^{r/2}}$, then we will split $P$ into $q = \ceil{ \frac{j - i}{D_r}}$ disjoint subpaths $P = P_1, P_2,...P_q$, such that every subpath except the last one has length $D_r$.  
This partition splits the path into at most $\lambda^rk^{r/2}$ subpaths where each subpath has length at most $D_r$, which is less than the length of the basic search distance in the level $r$.
Each related vertex to a path vertex $v_i$ in $G'$ is copied to $v_i$'s new logical path-relevant subproblem.
From \lemref{relatedvertices}, each subpath $P_i$ at level $r$ contains at most $2nk^{-r}$ related vertices.  We have at most $\lambda^r k^{c+r/2}$ new logical nodes since we have at most $\lambda^r k^{c+r/2}$ subpaths of length $\frac{D}{\lambda^r k^{c+r/2}}$. Hence, we only duplicate $2\lambda^rnk^{c-r/2}$ additional vertices in this procedure.
Next, we will construct the path-relevant subproblems tree based on the logical path-relevant subproblems in the following lemma, and show how to create the next level of subproblems from the logical subproblems layer.  Let $\rho_v$ be the scalar of the searching distance for pivot $v$.

\begin{lemma}
\lemlabel{core}
Consider a logical path-relevant subproblem $(G', P = \langle v_0, v_1,...,v_\ell \rangle, r)$ corresponding to a call to $\textsc{HSRecurse}(G', D, r)$. Let $p_r = \frac{\lambda k^{r+1}\log n}{n}$ be the probability a vertex is a pivot at level $r$.  
Let $S = \{v \mid \ell(v)=r, v \in R_{\rho_v D_r}(G', P)\}$ be the set of pivots at level $r$ related to $P$ within distance $\rho_v D_r$. There exists subpaths $P_0, P_1, P_2, ...,  P_{2\card{S}}$ such that,
\begin{enumerate}
\item If a vertex $v \in S$ is a $\rho_v D_r$-bridge, there are no path-relevant subproblems.
\item If no vertex $v \in S$ is a $\rho_v D_r$-bridge, then the vertex union of all $P_i$ for $0 \leq i \leq 2\card{S}$ is $P$.
\item $P_0, P_1, P_2, ...., P_{\card{S}+1}$ are in core problems and each $P_i$ is contained in some $V_{a_i}$.
\item $P_{\card{S}+1}$, ..., $P_{2\card{S}}$ are called in fringe problems and each $P_i$ is contained in some $V^{\textbf{fringe}}_{u}$, where $u \in S$.
\end{enumerate}
Additionally, with probability $1 - n^{-0.7\lambda + 4}$, we have that 
\begin{equation*} 
\begin{aligned}
\sum^{\card{S}}_{i=0}E[|R_{\rho_{min}D_{r}}(G'[V_{a_i}], P_i)|] \leq \frac{3}{ p_r}
\end{aligned}
\end{equation*}
and 
\begin{equation*} 
\begin{aligned}
\sum^{2|S|}_{i=|S|+1}E[|R_{\rho_{min}D_{r}}(G'[V^{\textbf{fringe}}_{u\in S}], P_i)|] \leq \frac{1}{p_r}
\end{aligned}
\end{equation*}
\end{lemma}

\begin{proof}
If $u$ is a pivot and $u \preceq_{\rho_u D_r} P$ or $P \preceq_{\rho_u D_r} u$, then $u$ will put a $u^{Anc}$ or $u^{Dec}$ label on vertices on path $P$. Otherwise, $u$ does not put a label on any vertices on $P$. If $u \preceq_{\rho_u D_r} P$ and $P \preceq_{\rho_u D_r} u$, then $u$ is a bridge and we will stop at this path-relevant subproblem. 

We now consider the case where all the pivots are ancestors. Let $u$ be an ancestor pivot, and we will show $u$ divides $P$ into three subpaths, which are contained in path-relevant subproblems.  
The first subpath is contained in the fringe problem, the second one is contained in the core problem with label $u^{Des}$, and the third one is in the core problem without a label from $u$. To define these subproblems, consider the indices of the following two nodes on $P$. The first index is the node that is the earliest $\rho_u D_r$-descendant to $u$ on $P$,
\begin{equation*} 
\begin{aligned}
Fringe(P, u) =  min\{i \mid  u \preceq_{{\rho_uD_r}} v_i\}.
\end{aligned}
\end{equation*}
 The second index is the node that is the earliest  $(\rho_u - 1) D_r$-descendant to $u$ on $P$,
\begin{equation*} 
\begin{aligned}
Core(P, u) = min(min\{j \mid  u \preceq_{{(\rho_u-1)D_r}} v_j\}, \ell + 1).
\end{aligned}
\end{equation*}
If no node on path $P$ is $(\rho_u - 1) D_r$-descendant of $u$, then we set $Core(P, u) = \ell + 1$. 
Since any $v_j$ which is a $(\rho_u - 1) D_r$-descendant to $u$ is also a $\rho_u D_r$-descendant to $u$, $Fringe(P, u) \leq Core(P, u)$.
Now we can split the path to three subpaths, $P^{Unrelated} = \langle v_0, ..., v_{Fringe(P, u) -1}\rangle$, $P^{Fringe} = \langle v_{Fringe(P, u)},..., v_{Core(P, u) - 1}\rangle$ and $P^{Core} = \langle v_{Core(P, u)},..., v_\ell \rangle$. 
If $Fringe(P, u) = 0$, then the unrelated subpath is defined as empty. 
If $Fringe(P, u) = Core(P, u)$, then the fringe subpath is defined as empty. 
If $Core(P, u) = \ell + 1$, then the core subpath is defined as empty.

Next we will show that $P^{Unrelated}, P^{Fringe}$ and $P^{Core}$ are contained in path-relevant subproblems. 
$P^{Unrelated}$ is trivial because $P^{Unrelated}$ doesn't get any labels from $u$. 
For any node $v_{i'} \in P^{Fringe}$, $u \not\preceq_{{(\rho_u-1)D_r}} v_{i'}$, because if $u \preceq_{{(\rho_u-1)D_r}} v_{i'}$, then $i' \geq Core(P, u)$, which contradicts the fact $i' < Core(P, u)$. 
On the other hand, we can show $u \preceq_{{(\rho_u + 1)D_r}} v_{i'}$, and so $v_{i'} \in R_{(\rho +1)D_r}(G,u) \backslash R_{(\rho -1)D_r}(G,u)$, which implies $P^{Fringe}$ is in $u$'s fringe problem.  
We have $u \preceq_{\rho_u D_r} v_{Fringe(P, u)}$ by definition, and $v_{Fringe(P, u)} \preceq_{D_r} v_{i'}$ because the length of $P$ is at most $D_r$.  
Therefore $u \preceq_{(\rho_u + 1) D_r} v_{i'}$.
Lastly, we will show that any node $v_{j'} \in P^{Core}$, is a $\rho_u D_r$-descendant of $u$ and therefore gets $u^{Des}$ label.  Since
$u \preceq_{(\rho_u - 1) D_r} v_{Core(P, u)}$ by definition, and $v_{Core(P, u)} \preceq_{D_r} v_{j'}$ by the length of the path being at most $D_r$, $u \preceq_{\rho_u D_r} v_{j'}$.
Thus, we have shown that each pivot $u$ will split $P$ into at most three subpaths, one that does not get a label from $u$, one that is in $u$'s fringe problem and one that gets $u^{Des}$ label.

Now we consider the case that there are $t$ ancestor pivots.  We will show the subpaths that are contained in the unrelated, fringe, and core problems, and that the union of these subpaths is the whole path $P$. 
For each pivot $u$, define $P^{Fringe}_u = \langle v_{Fringe(P, u)},..., v_{Core(P, u) - 1} \rangle$.  $P^{Fringe}_u$ is in $u$'s fringe problem. 
Consider all $Core(P, u)$ values for all $t$ ancestor pivots in non-decreasing order, $c_1, c_2, ..., c_{t}$. 
Let $u_1, ..., u_i, ..., u_{t}$ be the corresponding ancestor pivots and for convenience, set $c_0 = 0$ and $c_{t+1} = \ell + 1$. 
Each vertex on the path $\langle v_{c_{i}},..., v_{c_{i+1}-1} \rangle$ gets a label from each of the $u_1,..., u_i$ pivots.
However, the subpaths become more complicated because of the fringe problem. In particular, if a node is in a fringe problem, we no longer need to consider it in the core problem. 
The $i$-th core path is $P_i = \langle v_{c_{i}},..., v_{f_i - 1}\rangle$, where $f_i$ is defined as $f_i = min(min\{Fringe(P, u_j) \mid j > i\}, c_{i+1})$. 
$P_i$ gets a label from each $u_1, ..., u_{i}$ because $c_i \geq Core(P, u_{j})$ for all $j \leq i$. 
On the other hand, the end index of $P_i$ is at most $f_{i} - 1$, which means $P_i$ does not get a label from any of $u_{i+1}, ..., u_{t}$. Therefore, each $P_i$ will be in a core problem. Finally, define $P_0 = \langle v_0,..., v_{c_{1}-1} \rangle$ and $P_{i} = P^{Core}_{u_i}$ and so there are at most $t + 1$ core problems.

Now we will show the union of these subpaths is the whole path $P$.
We have showed any vertices in $P_0$ and $P_i$, $i \in [1,t]$ are in the unrelated and core problems, respectively.  The remaining vertices on $P$ are vertices in $\langle v_{f_i}, ..., v_{c_{i+1}-1} \rangle$.  By definition, these are in fringe problems.  Therefore, the union of $P_0$, $P_u^{Fringe}$, and $P_u^{Core}$ for $i \in [1,t]$ is the path $P$.

Next, we need to consider the case where some pivots are descendants. This case becomes more complicated, but the basic idea is the same as just ancestor pivots. 
We will define the $P^{Core}$ and $P^{Fringe}$ subpaths and show that each vertex on these subpaths gets the appropriate label.  Then we will show the union of these subpaths is the entire path.
We first need to define $Fringe(P,u)$ and $Core(P,u)$ for a descendant pivot $u$.  $Fringe(P,u)$ is the index of the latest path node that is a $\rho_u D_r$ ancestor of $u$, 
\begin{equation*} 
\begin{aligned}
Fringe(P, u) =  max\{i \mid  v_i \preceq_{{\rho_u D_r}} u\} + 1.
\end{aligned}
\end{equation*}
 The core node is the latest node that is $(\rho_u - 1) D_r$-ancestor of $u$ on $P$,
\begin{equation*} 
\begin{aligned}
Core(P, u) = max(max\{j \mid  v_j \preceq_{{(\rho_u-1)D_r}} u\}, -1) + 1.
\end{aligned}
\end{equation*}
For our convenience, we shift $Fringe(P, u)$ and $Core(P, u)$ by 1 index. 
If no node on path $P$ is a $(\rho_u - 1) D_r$-ancestor of $u$, then set $Core(P, u) = 0$.  We use the same strategy as the ancestor case to define all subpaths $P^{Fringe}$ in the fringe problem. 
If $u$ is a descendant pivot, then $P^{Fringe}_u = \langle v_{Core(P, u)}, ..., v_{Fringe(P, u) - 1}  \rangle$, and if $u$ is an ancestor pivot, $P^{Fringe}_u = \langle v_{Fringe(P, u)}, ..., v_{Core(P, u) - 1} \rangle$. It is clear that $P^{Fringe}_u$ is in $u$'s fringe problem based on the ancestor case. 
For the core problems, consider the $\card{S}$ $Core(P, u)$ values in non-decreasing order, $c_1, c_2, ..., c_{|S|}$. 
Let $u_1, ..., u_i,..., u_{\card{S}}$ be the corresponding pivot, and for convenience, set $c_0 = 0$ and $c_{|S| + 1} = \ell$.
Notice that some of the $u_i$ are ancestor pivots, and some are descendant pivots. 
If a path node $v_i$ is not in a fringe problem, then $v_i$ will get a $u_j^{Des}$ label from each ancestor pivot $u_j$ where $c_j \leq i$ and a $u_{j'}^{Anc}$ label from each descendant pivot $u_{j'}$ where $c_{j'} \geq i$.
We will next show the subpaths in more detail.
First define $f_i = min( min\{ Fringe(P, u_j) \mid j > i\}, c_{i + 1})$ and $g_{i} = max(max\{Fringe(P, u_j) \mid j \leq i\}, c_i)$.
Now we can define path $P_i = \langle v_{g_{i}}, ..., v_{f_{i} - 1} \rangle$.  It's possible that $g_i \geq f_i$ in which case $P_i$ is empty.
Let $A_i = \{ u_j \mid j \leq i, u_j \textmd{ is an ancestor pivot} \}$ be the set of ancestors on the "left" of $u_i$, and $\Bar{A_{i}} = \{u_j \mid j > i, u_j \textmd{ is an ancestor pivot}\}$ be the set of ancestors on the "right" of $u_i$. We want to show each node on $P_i$ gets a $u_j^{Des}$ label from each $u_j \in A_i$ and does not get any labels from nodes in $\Bar{A_{i}}$. 
The first claim is trivial since the core value of the nodes in $A_i$ is less than or equal to $c_i$, and so $P_i$ gets a label from each node in $A_i$. 
For the second claim, if any node $u'$ on $P_i$ gets label from ancestor pivot $u_j \in \Bar{A_i}$, then $Fringe(P, u_j) \leq f_i - 1$ in order for the search from $u_j$ to reach a node on $P_i$. 
However, this is a contradiction the definition of $f_i$.  Therefore $P_i$ does not get any labels from $u_j \in \bar{A_i}$.
Next we will show the descendants case.  Define $D_i = \{ u_j \mid j \geq i + 1, u_j \textmd{ is a descendant pivot \}}$ and $\bar{D_i} = \{ u_j \mid j < i + 1, u_j \textmd{ is a descendant pivot \}}$.  These definitions are symmetric to the case for ancestors except they are shifted by 1 index to account for the shift by 1 in the definition in $Fringe(P,u)$ and $Core(P,u)$ for descendants.  The proof that all nodes in $P_i = \langle v_{g_i}, ..., v_{f_i-1} \rangle$ get an $Anc$ label from each vertex in $D_i$, and get no labels from any vertex in $\bar{D_i}$ is symmetric to the ancestor case with the exception of the shift by 1 index.  We can conclude that each $P_i$ will be in a core problem.  There are $P_0, P_1, ..., P_{\card{S}}$ subpaths, which is at most $\card{S} + 1$ subpaths.

The last thing left to prove is that the union of $P_i$ for $i \in [1,\card{S}]$, and $P^{Fringe}_u$ for all $u \in S$ is $P$. This claim is straightforward because 
$\langle v_{g_i}, ..., v_{f_i} \rangle$, for $i \in [1,\card{S}]$, are in core problems, and $\langle v_{c_i}, ..., v_{g_i - 1}\rangle$ and $\langle v_{f_i}, ..., v_{c_{i + 1} - 1} \rangle$ are in fringe problems based on their definitions.

Now we will show the last part of the claim, which is the number of related nodes of $P$ at level $r+1$.  First, we will count the related nodes in the core problems and then the fringe problems.  
We will slightly overcount the related nodes in the core problems by counting all core path-relevant subproblems created by $\rho_{min} D_r$-pivots. Notice that all $\rho_{min} D_r$-pivots will be in $S$ because the searching distance is at least $\rho_{min} + 1$. If we only consider part of $S$ as pivots, we will increase the related nodes at level $r+1$.

Consider the related nodes in the core problem at level $r+1$.  Any ancestor pivot that reaches the path in $\rho_{min} D_r$, reaches $v_{\ell}$ in $\rho_{max} D_r$ because the length of $P$ is at most $D_r$.  Using this fact, and \lemref{relatedvertices}, with probability $1 - n^{-0.7\lambda + 3}$, the following holds,
\begin{equation*} 
\begin{aligned}
|R^{-}_{\rho_{min}D_{r}}(G, P)| \leq |R^{-}_{\rho_{max}D_{r}}(G, v_\ell)| \leq nk^{-r}.
\end{aligned}
\end{equation*}
Assume that $|R^{-}_{\rho_{min}D_{r}}(G, P)| > 1.5 / p_r$, otherwise the claim holds trivially.
Let $Y_s$ be the number of ancestors at level $r+1$ in a logical path-relevant subproblem $s$, and $X_s$ be the $\rho_{min} D_r$ ancestors pivots in logical path-relevant subproblem $s$.
Note $X_s$ is a binomial random variable of $B(|R^{-}_{\rho_{min}D_{r}}(G, P)|,p_r)$, and $E[X_s] = |R^{-}_{\rho_{min}D_{r}}(G, P)| \cdot p_r \leq \lambda k \log n$ with probability $1 - n^{-0.7\lambda + 3}$, where the probability comes from \lemref{relatedvertices}. 
Using a Chernoff bound, the number of ancestor pivots is bounded with high probability as follows,
\begin{equation*}  
\Prob{X_s \geq 4\lambda k \log n} \leq e^{-2\lambda k n \log n} \leq n^{-2\lambda}.
\end{equation*}
Hence, by taking a union bound over all $v \in V$ and all $r$, the rest of the proof holds with probability $1 - n^{-0.7\lambda + 4}$.

Notice that the scalar interval $|I| = 4\lambda^2 k \log^2 n$.  If $\lambda \geq 4$, then $\card{I} \geq 4X_s$. 
In \lemref{relateddecreases} we will show that if $\card{I} \geq 4X_s$ then,
\begin{equation*}  
E[Y_s \mid X_s] \leq \frac{1.5}{X_s + 1} \cdot |R^{-}_{\rho_{min}D_{r}}(G, P)|.
\end{equation*}
The expectation of $Y_s$ is,
\begin{equation*} 
\begin{aligned}
E[Y_s ] &= \sum_s E[Y_s \mid X_s] \cdot \Pr[X_s] \leq  1.5|R^{-}_{\rho_{min}D_{r}}(G, P)| \cdot \sum_s \frac{1}{X_s + 1} \Pr[X_s] \\
&\leq 1.5|R^{-}_{\rho_{min}D_{r}}(G, P)| \cdot E[\frac{1}{X_s + 1}] \\ &\leq 1.5|R^{-}_{\rho_{min}D_{r}}(G, P)| \cdot \frac{1}{|R^{-}_{\rho_{min}D_{r}}(G, P)| \cdot p_r} = \frac{1.5}{p_r} ,
\end{aligned}
\end{equation*}
where the first line comes from the definition of expectation and \lemref{relateddecreases}, the second line comes from the definition of expectation, and the last line comes from $X_s$ being a binomial random variable.
The descendants case is symmetric. 
In total, 
\begin{equation*} 
\begin{aligned}
\sum^{|S|}_{i=0}E[|R_{\rho_{min} D_r}(G'[V_{a_i}], P_i)|] \leq \frac{3}{ p_r} ,
\end{aligned}
\end{equation*}
with probability $1 - n^{-0.7\lambda + 4}$.

Lastly, we will count the related nodes in the fringe problems. 
Notice that for all pivots $u$, $u \in R_{(\rho_{max}-1)D_{r}}(G, P)$, and earlier in the proof we showed $\card{R_{(\rho_{max}-1)D_{r}}(G, P)} \leq \card{R^{+}_{\rho_{max}D_{r}}(G, v_0)} + \card{R^{-}_{\rho_{max}D_{r}}(G, v_\ell)} \leq 2nk^{-r}$. 
By \lemref{fringevertices}, each node adds $1/(4\lambda k \log n)$ nodes to its fringe problem in expectation. Therefore,  
\begin{equation*} 
\begin{aligned}
\sum^{2|S|}_{i=|S| + 1}E[|R_{\rho_{min} D_r}(G'[V^{fringe}_{u}], P_i)|] \leq \card{R_{(\rho_{max}-1)D_{r}}(G, P)} \cdot \frac{1}{4 \lambda k \log n} \leq \frac{2}{4 \lambda k^{r+1} \log n/ n} \leq  \frac{1}{p_r}.
\end{aligned}
\end{equation*}
\end{proof}

Now we will show the helper lemma for the case the pivots are ancestors and descendants.
Fineman \cite{Fineman18} shows a similar result when there is just one pivot, and JLS \cite{Stanford19} extends this to $t$ pivots.  In our case, we have the additional difficulty that each pivot searches for a different distance, but we are able to get the same result.

\begin{lemma}
\lemlabel{relateddecreases}
Consider the path $P = \langle v_0, v_1, .., v_{\ell} \rangle$, where $\ell \leq D_r$, and its $\rho_{min}D_r$-distance ancestor set $R^{-}_{\rho_{min}D_r}(G, P)$ in the $r^{th}$ level of recursion.
Let $I$ be the set containing all possible values of interval scalar.
Choose $t$ ancestor pivots uniformly at random from $R^{-}_{\rho_{min}D_r}(G, P)$. 
Let $P_i$ be the path defined in \lemref{core}. 
If the chosen interval $\card{I} \geq 4t$, then
\begin{equation*} 
\begin{aligned}
\sum^{|S|}_{i=0}E[ |R^{-}_{\rho_{min}D_{r}}(G'[V_{a_i}], P_i)|] \leq \frac{1.5}{t + 1}\card{R^{-}_{\rho_{min}D_{r}}(G', P)}. 
\end{aligned}
\end{equation*}
\end{lemma}

\begin{proof}
For two nodes, $u,w$, define the relation $u$ \textbf{knocks out} $w$ to mean that if $u$ is a pivot and $w$ is not in level $r$, then $w$ is not path-relevant in a path-relevant subproblem in the level $r+1$ i.e. $w \notin R^{-}_{\rho_{min}D_{r}}(G'[V_{a_i}], P_i)$ for all $i \in [1, \card{S}]$. 
If $u$ does not knock out $w$, it does not mean $w$ is path-relevant in a path-relevant subproblem in the next level. Other pivots may knock out $w$, in which case $w$ would not be path-relevant. 
In the proof, we will prove our lemma by three steps. In the beginning, recall $u$ and $w$ will choose an interval scalars $\sigma_u$ and $\sigma_v$ if they are chosen as pivots, respectively. We will show that for most interval scalar in $I$, either $u$ knocks out $v$ or $v$ knocks out $u$. Secondly, based on the first claim, we are able to show for any fixed set $S' \in R^{-}_{\rho_{min}D_{r}}(G', P), |S'| = t+1$, if we choose $t$ pivots in $S'$, the left node will be a $\rho_{min}D_{r}$-ancestor in next level with probability $1.5/(t + 1)$. Last, we need to use the second claim to prove if we choose $t$ pivots from $R^{-}_{\rho_{min}D_{r}}(G', P)$, the expectation in the lemma claim holds. 

For any two vertices $u, w$ and interval scalar $\sigma \in I$, let $\chi(u, w, \sigma)$ be the indicator that $u$ doesn't knocks out $w$, when $\sigma$ is the interval scalar for $u$.
We will show that for $u$ and $w$, there is at most one scalar $\sigma \in I$ where $\chi(u, w, \sigma) = 1$ and $\chi(w, u, \sigma) = 1$. 
W.l.o.g. assume $\id{dist}_{G'}(u,w) \leq \id{dist}_{G'}(w,u)$. Let $h$ be $\id{dist}_{G'}(u,w)$. Recall that the algorithm chooses a random interval defined by the interval scalar $\sigma$, and then chooses another scalar $\rho$ from the corresponding interval, which minimizes the size of the fringe problem. As a slight of abuse of notation, let the interval with corresponding $\sigma$ be denoted $[\sigma_{min}, \sigma_{max})$. Therefore, $\rho \in [\sigma_{min}, \sigma_{max})$.
First, we will show that in the cases where $\sigma_{min}D_r > h$ and $\sigma_{max}D_r \leq h$, at most one of $\chi(u, w, \sigma)$ and $\chi(w, u, \sigma)$ equals $0$.  Therefore, only the interval $[\sigma_{min}, \sigma_{max})$ can make $\chi(u, w, \sigma) = 1$ and $\chi(w, u, \sigma) = 1$.

For the first case where $h < \sigma_{min}D_r$, Fineman \cite{Fineman18} has given a similar proof that either $u$ knocks out $w$ or $w$ knocks out $u$. 
If $w$ is a pivot, then $u$ gets a $w^{Anc}$ because $u \preceq_{\rho_uD_r} w$, where $\rho_u$ is the scalar for $u$ and $\rho_u \geq \sigma_{min}$. 
However, in the proof of \lemref{core}, we showed that none of subpath $P_{i}$ in the core problem gets $w^{Anc}$ label. Therefore, $w$ knocks out $u$.

The second case is $h \geq \sigma_{max}D_r$.  
Let $c_u = Core(P, u)$ and $c_w = Core(P, w)$. 
Notice that for each $\rho_{min}D_r$-distance ancestor pivot, since the searching distance is at least $(\rho_{min} + 1) D_r$, $c_u \leq \ell$ and $c_w \leq \ell$. 
Consider two more indices, $c'_u = min\{j \mid u \preceq_{\rho_{min}D_r} v_j\}$, and $c'_w = min\{j \mid w \preceq_{\rho_{min}D_r} v_j\}$. 
Notice that $u, w \in R^{-}_{\rho_{min}D_r}(G, P)$, and so $c'_u$ and $c'_w$ are always valid. 
Even if $u$ is not pivot, $u$ can only be a $\rho_{min}D_r$-ancestor for a subpath $P_{i}$ such that $\id{tail}(P_i) \geq c'_u$. 
This is because the shortest path between two nodes never decreases in the algorithm. If $u$ is not a $\rho_{min}D_r$-ancestor for any subpath $P'$, $u \notin R^{-}_{\rho_{min}D_{r}}(G'[V'], P')$ holds throughout the rest of the algorithm. 
Note that $c_u \leq c'_u$ and $c_w \leq c'_w$, because $\rho_u \geq \rho_{min} + 1$ and $u \preceq_{(\rho_{u} - 1)D_r} v_{c'_u}$ based on the definition of $c'_u$. 
First assume $c_u \leq c'_w$.  If $u$ is a pivot, and the subpath for $u$ is $P_{i_u}$, in \lemref{core} we have shown all paths $P_i$ get $u$'s descendant label, where $i \geq i_u$. 
However, $w$ never gets a descendant label from $u$ since $h \geq \sigma_{max}D_r > \rho_u D_r$. Thus, $w$ could not be a  $\rho_{min}D_{r}$-ancestor for all paths $P_j$ where $j \geq i_u$.
On the other hand, $w$ could not be a  $\rho_{min}D_{r}$-ancestor for all paths $P_j$ where $j < i_u$, since $\id{tail}(P_j) < c_u \leq c'_w$ and we already know $w$ could not be a $\rho_{min}D_r$-ancestor for subpath $P_{j}$ if $\id{tail}(P_j) < c'_w$. 
Therefore, $\chi(u, w, \sigma) = 0$ if $c_u \leq c'_w$.
 If $c_w \leq c'_u$, the situation will be symmetric. 
In order for $\chi(u, w, \sigma) = 1$ and $\chi(w, u, \sigma) = 1$ to hold at the same time for $h \geq \sigma_{max}D_r$, it has to be the case that $c_u > c'_w$ and $c_w > c'_u$. Combining with the fact $c_u \leq c'_u$ and $c_w \leq c'_w$, this is impossible. Therefore, when $h > \sigma_{max}D_r$, one of $\chi(w, u, \sigma)$ and $\chi(u, w, \sigma)$ is $0$.

Next, for any $t+1$ ancestors set $S'\in R^{-}_{\rho_{min}D_{r}}(G', P)$, Our goal is to show that if we randomly choose $t$ pivots from $S'$, the leftover node $u_i$ is path-relevant in the next level with probability at most $1.5/(t+1)$. For our convenience, denote $R^{-}_{\rho_{min}D_{r}}(G', P)$ as $R^{-}$.

Let $E_{left}$ be the event that left node $u_i$ is path-relevant in the next level. Notice that the probability is 
\begin{equation*} 
\begin{aligned}
\Prob{E_{left}}  = \frac{\sum^{t+1}_{i = 1}\sum_{\sigma_{u_1}, \sigma_{u_2},...,\sigma_{u_{i-1}},\sigma_{u_{i+1}},..,\sigma_{u_{t+1}} \in I^{\times t}}\prod_{j\neq i}\chi(u_j, u_i, \sigma_{u_j})}{(t+1) |I|^{t}}
\end{aligned}
\end{equation*}
That's because if $u_i$ is the left pivot and each $u_j$'s scalar is $\sigma_{u_j}$, where $j \neq i$, then $u_i$ is path-relevant in the next level if an only if all $u_j$ doesn't knock out $u_i$ for $j \neq i$, i.e., $\chi(u_j, u_i, \sigma_{u_j}) = 1$ holds for all $j \neq i$. 
Let $A_{ji} = |\{\sigma \mid \chi(u_j, u_i, \sigma) = 1, \sigma \in I \}|$ be the number of interval scalar such that $u_j$ doesn't knock out $u_i$, then we can rewrite the probability as 
\begin{equation*} 
\begin{aligned}
\Prob{E_{left}}  = \frac{\sum^{t+1}_{i = 1} \prod_{j \neq i}A_{ji}}{(t+1) |I|^{t}}
\end{aligned}
\end{equation*}
Notice that since for each $u_i$ and $u_j$, there is at most one $\sigma$ making $\chi(u_j, u_i, \sigma) = 1$ and $\chi(u_i, u_j, \sigma) = 1$. 
Therefore we have $A_{ij} + A_{ji} \leq |I| + 1$. 
$\sum^{t}_{i = 1} \prod_{j \neq i}A_{ji}$ will be maximized when one of $A_{ij}$ and $A_{ji}$ equals to $|I| + 1$. Therefore, 
\begin{equation*} 
\begin{aligned}
\Prob{E_{left}} \leq \frac{(|I| + 1)^t}{(t+1) |I|^{t}} \leq \frac{1}{t+1} \cdot (1 + \frac{1}{|I|})^{|I|/4} \leq \frac{1.5}{t + 1}. 
\end{aligned}
\end{equation*}
Last, we want to use $\Prob{E_{left}}$ show the expectation when we choose $t$ pivots in $R^{-}$. Let $\mathbb{I}(S, u)$ be the indicator that a node $u$ is path-relevant in the next level when $S$ is the chosen pivots. Thus, for any fixed $S'$, we have 
\begin{equation*} 
\begin{aligned}
\sum_{S \subset S', |S| = t} \mathbb{I}(S, {S' \backslash S}) = (t + 1)\Prob{E_{left}} \leq 1.5
\end{aligned}
\end{equation*}
Summing up over all possible $S' \in R^{-}$, 
\begin{equation*} 
\begin{aligned}
\sum_{\substack{S' \subset R^{-}, |S'| = t + 1\\S \subset S', |S| = t}} \mathbb{I}(S, {S' \backslash S}) \leq 1.5 {|R^{-}|\choose{t+1}}.
\end{aligned}
\end{equation*}
we can rewrite the formula as
\begin{equation*} 
\begin{aligned}
\sum_{\substack{S \subset R^{-}, |S| = t\\u \in R^{-} \backslash S, S' = S \cup \{u\}}} \mathbb{I}(S, {S' \backslash S}) \leq 1.5 {|R^{-}|\choose{t+1}}.
\end{aligned}
\end{equation*}
The above formula first chooses $S$ then choose a vertex $u \notin S$ while the previous formula first chooses $S'$ then chooses a vertex $u \in S'$. Notice that if $u \in S$, then $\mathbb{I}(S, u) = 0$. Therefore, 
\begin{equation*} 
\begin{aligned}
\sum_{\substack{S \subset R^{-}, |S| = t\\u \in R^{-}}} \mathbb{I}(S, u) \leq \sum_{\substack{S \subset R^{-}, |S| = t\\u \in R^{-} \backslash S, S' = S \cup \{u\}}} \mathbb{I}(S, {S' \backslash S}) \leq 1.5 {|R^{-}|\choose{t+1}}.
\end{aligned}
\end{equation*}

Hence, if the algorithm chooses $t$ ancestor pivots, the expected number of ancestors in the next level is 
\begin{equation*} 
\begin{aligned}
&\sum^{|S|+1}_{i=1}E_{|S| = t}[ |R^{-}_{\rho_{min} D_r}(G'[V_{a_i}], P_i)| ] = E_{|S| = t}[\sum_{u \in R^{-}}\mathbb{I}(S, u)] \\
= &\sum_{\substack{S \subset R^{-}, |S| = t\\u \in R^{-}}} \mathbb{I}(S, u) \Prob{S} \leq 1.5 {|R^{-}|\choose{t+1}} / {|R^{-}|\choose{t}} \\ 
\leq &\frac{1.5}{t + 1}|R^{-}_{\rho_{min}D_{i}}(G', P)|.
\end{aligned}
\end{equation*}
\end{proof}

Notice that each subpath $P_i$ will be contained in a subproblem, which means all $P_i$ are valid in problems even if they were split in the logic layer. There might be some path-relevant subproblems replicated multiple times, so these path-relevant subproblems are no longer independent. 
Another thing is each subpath length $|P_i| \leq \frac{D}{\lambda^r k^{r/2}}$, and we will construct new logical layer based the rule we mentioned before. 
Next, based on the path-relevant subproblems tree, we will give a theorem about the expected number of related nodes and subproblems in each level.

\begin{lemma}
\lemlabel{subproblemsize}
Consider the path-relevant subproblems tree for one execution of \textsc{Hopset}$(G)$.
 Let $Z_r$ be the number of subproblems in the $r^{th}$ level of recursion.  For all $r \geq 0$, 
 \begin{align*}
\bigcap_{r \leq \log_k n - L} \Prob{  Z_{r} \leq 32\lambda^{r}  k^{c + \frac{r + 1}{2}} \log^2 n} \geq \frac{1}{2}.
 \end{align*}

\end{lemma}

\begin{proof}
To show the claim, we will first show the expectation of $Z_r$. Let $Y_r$ be the number of path related vertices in the $r^{th}$ level of recursion. Our target is the following formula holds with probability $1 - n^{-0.7\lambda + 4}$, for all $r$,
\begin{align*}
E[Y_{r}] &\leq 4\lambda^{r} n k^{c-\frac{r}{2}} \\  
E[Z_{r}] &\leq 15\lambda^{r}  k^{c + \frac{r + 1}{2}} \log n.
\end{align*}
If the expectation of $Z_r$ in the above formula holds, then by Markov's inequality, 
\begin{align*}
\Prob{Z_{r} \geq 30\lambda^{r}  k^{c + \frac{r + 1}{2}} \log^2 n} \leq \frac{1}{2\log n},
\end{align*}
and by union bound, the following holds if $\lambda \geq 8$,
\begin{align*}
\bigcup_{r \leq log_k n - L} \Prob{ Z_{r} \geq 32\lambda^{r}  k^{c + \frac{r + 1}{2}} \log^2 n} \leq \frac{\log_k n - L}{2\log n} + n^{-0.7\lambda + 4} \leq \frac{1}{2}.
\end{align*}
Next we will show the expectation of $Z_r$ and $Y_r$ by induction on the level of recursion.  When $r=0$, the claim is trivial since there is one subproblem and at most $n$ path-related vertices.
Assume for level $r$, the formulas hold. Then we will construct the logical layer.
Let $Y'_r$ be the number of path-related nodes in the logical layer at level $r$.  Let $Z'_r$ be the number of subproblems in the logical layer at level $r$. The search distance for level $r$ is $D/\lambda^rk^{r/2}$ and subproblem is duplicated if the path length in the subproblem is greater than $\frac{\ell}{\lambda^rk^{c+r/2}}$. Thus, at most $\lambda^rk^{c+r/2}$ subproblems are duplicated and $Z'_r = Z_r + \lambda^{r} k^{c+r/2}$.
On the other side, from \lemref{relatedvertices}, the number of related nodes in each subproblem at level $r$ is less than or equal to $2 nk^{-r} $ with probability $1 - n^{-0.7\lambda + 3}$. Hence,
\begin{equation*}  
\begin{aligned}
Y'_r = Y_r + \lambda^{r}k^{c+r/2} \cdot 2 nk^{-r} = Y_r + 2\lambda^{r}n k^{c-r/2}
\end{aligned}
\end{equation*}
Next we can count the $Z_{r+1}$ and $Y_{r + 1}$ based on the logical layer. By \lemref{core}, for each subproblem at level $r$, the number of related nodes at level $r+1$ can be bounded. For a logical subproblem $s$ at level $r$, let $Y_s$ be the number of path-related nodes in $s$'s subproblem at level $r + 1$.  The expectation of $Z_{r+1}$ is,

\begin{equation*}  
\begin{aligned}
E[Y_{r+1}] =& E[ \sum_s Y_s] = \sum_s E[Y_s] = \sum_{Z'_r}\sum_s E[Y_s \mid Z'_r] \Pr[Z'_r] \\
\leq & \frac{4}{p_r} \sum_{Z'_r} Z'_r \Pr[Z'_r] = \frac{4}{\lambda k^{r+1}\log n / n} \cdot (E[Z_r] + \lambda^{r}k^{c+r/2}) \\
\leq &  64 \lambda^{r - 1}n k^{c-\frac{r+1}{2}} \leq 4 \lambda^{r + 1}n k^{c-\frac{r+1}{2}}
\end{aligned}
\end{equation*}
for $\lambda \geq 4$. For the $Z_{r+1}$, if there are $t$ pivots, there will be at most $2t + 1$ subproblems. To count $Z_{r+1}$, split $2t+1$ subproblems to two parts, $2t$ subproblems and $1$ subproblem. The $2t$ part will contribute to the total number of pivots. On the other hand, each subproblem at level $r$ will have 1 additional subproblem, which implies another $Z'_{r}$ item. Therefore, if $k \geq 2$ then,
\begin{equation*}  
\begin{aligned}
E[Z_{r+1}] =& \sum E[ Z_{r+1} \mid Y'_{r}] \cdot \Pr[Y'_{r}]= p_r \cdot \sum 2Y'_{r} \Pr[Y^{'}_{r}] + E[Z'_{r}]\\
=& \frac{2\lambda k^{r+1}\log n}{n} \cdot E[Y'_{r}] + E[Z'_{r}] \\
\leq& \frac{2\lambda k^{r+1}\log n}{n} \cdot (4\lambda^rnk^{c-r/2} + 2\lambda^rnk^{c-r/2}) + 15\lambda^rk^{c+\frac{r+1}{2}}\log n + \lambda^rk^{c+r/2} \\
\leq&  15 \lambda^{r+1} k^{c + 1 + r / 2} \log n.
\end{aligned}
\end{equation*}
\end{proof}
Lastly, we will show the hopbound based on the path-relevant subproblems tree.

\begin{lemma} \lemlabel{hopboundedges}
Consider any graph $G = (V,E)$ and any shortest path $|\hat{P}| \geq n^{1/2}$ from $u$ to $v$. Consider an execution of Algorithm \ref{hopset}. Let $E'$ be the hopset produced, and let $Z_0, Z_1,..., Z_{r}$ be the number of corresponding path-relevant tree subproblems at level $r$, then there is a $u$-to-$v$ path in $G' =(V,E' \cup E)$ containing at most $3\sum_{r \leq \log_kn - L}Z_r$ edges.
\end{lemma}
\begin{proof}
A path-relevant subproblems tree node will have no children if the subproblem contains a path-relevant pivot that is a bridge.  
If any pivots $w$, are bridges at or before level $L$, then $w$ will be a shortcutter in Algorithm \ref{hopset}. 
Notice that $w$ is $\rho_{max}D_r$-related to $\hat{P}$ for $r \leq L$. We require that $\rho_{max}D_0 \leq \ell$ since we only search for additional $\ell$ distance. The new path will be $u$ to $w$ to $v$. 

Otherwise, there are no bridges in the first $L$ levels. 
Consider a path-relevant subproblem at level $r' > L$.  
If there is a pivot $w$ at level $r'$ that is a bridge, then at level $r'-L$ $w$ was a shortcutter in a path-relevant subproblem $(G,P',r'-L)$.
In \lemref{core} we have shown that $P' \leq D_r$.  
Since shortcutters search for $\rho_{max}D_r$, $w$ reaches $\id{head}(P')$ and $\id{tail}(P')$, $(\id{head}(P'),w)$ and $(w, \id{tail}(P'))$ are added to $E'$, creating a two hop path from $u$ to $v$ in $G'$.
At level $\log_k n$, all vertices are pivots, and therefore the path must have a bridge pivot.
In total there are at most $2\sum_{r \leq \log_kn - L}Z_r$ hopset edges that shortcut path-relevant subproblems, and there are at most $\sum_{r \leq \log_kn - L}Z_r$ edges between subproblems.  Adding these together completes the proof.
\end{proof}

\begin{lemma}
Consider any graph $G' = (V,E)$ and an execution of \textsc{Hopset}$(G')$ with parameters $k$, $\lambda$ and $L$.  The hopset produced has hopbound $n^ {\frac{1}{2} + O(1/\log k)} k^{c + \frac{1-L}{2}} \log^2 n$ with probability $1 - n^{-\lambda + 2}$.
\end{lemma}

\begin{proof}
Consider any shortest path $\hat{P}$ with $|\hat{P}| > n^{1/2}$ from $u$ to $v$.  By \lemref{hopboundedges}, there is a path from $u$ to $v$ with at most $3\sum_{r \leq \log_kn - L}Z_r$ edges where $Z_r$ is the number of path-relevant subproblems in the path-relevant subproblems tree at level $r$.
Since the algorithm is repeated $\lambda \log n$ times, there exists a path relevant tree such that $\bigcap_{r \leq \log_k n - L} Z_{r} \leq 32\lambda^{r}  k^{c + \frac{r + 1}{2}} \log^2 n$ holds with probability $1 - n^{-\lambda}$, by  \lemref{subproblemsize}.
Therefore the hopbound is, 
\begin{equation*}  
\begin{aligned}
\sum_{r \leq \log_kn - L} 3Z_r = \sum_{r \leq \log_kn - L} 96\lambda^r k^{c + \frac{r + 1}{2}} \log^2 n  = n ^{\frac{1}{2} + O(1/\log k)}k^{c + \frac{1-L}{2}} \log^2 n,
\end{aligned}
\end{equation*}
with probability $1 - n^{-\lambda + 2}$, where the probability comes from taking a union bound over all possible shortest paths.
\end{proof}
\subsection{Approximation}
\label{section:approximation}
We have already showed that the path-relevant tree has $n^ {\frac{1}{2} + O(1/\log k)} k^{c + \frac{1-L}{2}} \log^2 n$ nodes, which means there exist a path $P'$ that contains at most $n^ {\frac{1}{2} + O(1/\log k)} k^{c + \frac{1-L}{2}} \log^2 n$ hops. Now we want to show that $P'$ is an good approximation of the original path $\hat{P}$. 
Notice that in the path-relevant tree, a path-relevant problem has no subproblems if one of the pivots at that level is a bridge.  Consider the following two cases:
\begin{enumerate}
     \item If there is a bridge $u$ with $\ell(u) \leq L$, then we stop the path-relevant tree at level 0. In this case, the searching distance is at most $D \in [ \ell k^{-c},  2\ell k^{-c})$, so the bridge will have at most $2 \cdot 32 \lambda^2  k^2 \log^2 n \cdot D \leq 128 \lambda^2 k^{2 - c} \log^2 n \cdot \ell$ error. The 2 comes from the forward and backward searches, the second item $32\lambda^2  k^2 \log^2 n$ comes from the scaling factor.
    \item Consider the path-relevant tree after level 0.  If a path-relevant subproblem selects a shortcutter that is a bridge at level $r + L$, then the path-relevant subproblem will end at level $r$. The error for this subproblem is at level $r$ is at most $2 \cdot 32 \lambda^2 k^2 \log^2 n \cdot D_{r+L}$. Summing up all possible briges, we have the error 
    \begin{align*}
       \sum^{r = \log_k n - L}_{r = 1} Z_r \cdot 64 \lambda^2 k^2 \log^2 n \cdot D_{r+L} \leq 4096\lambda^{2-L}k^{(5-L)/2}\log^5 n \cdot \ell
    \end{align*}
    Hence, the accumulating error will be $4096\lambda^{2-L}k^{(5-L)/2}\log^5 n \cdot \ell$.
\end{enumerate}

To make the first error equal to second error, set $k^c = \frac{\lambda^Lk^{(L-1)/2}}{32\log^3 n}$. If $k = \Omega(\log n)$ and the desired error is $\epsilon \ell$, set $L = 15 - 2\log_k \epsilon$. The hopbound $\beta$ is at most $6\lambda^{\log_kn}n^{1/2} / \log n$. The running time is $O(mk^{16}\log^4n /\epsilon^2)$ and the hopset size is $O(nk^{16}\log^4n /\epsilon^2)$. 
Combining all this together, the following corollary holds.

\begin{corollary}
\label{unweightedhopbounds}
For any unweighted directed graph $G = (V,E)$, \textsc{Hopset}$(G)$ with above parameter returns a $(\beta = n^{1/2+O(1/\log k)}, \epsilon)$-hopset of size $O(nk^{16}\log^4 n /\epsilon^2)$ in running time $O(mk^{16}\log^4n /\epsilon^2)$  with probability $1 - n^{-\lambda + 2}$.
\end{corollary}

\begin{proof}[Proof of Theorem \ref{unweightedmaintheorem}.]
From Theorem \ref{runtimetheorem} and Corollary \ref{unweightedhopbounds}, Theorem \ref{unweightedmaintheorem} follows directly. 
\end{proof}

\section{Weighted Graphs} \label{section:weighted}

In this section, we present an algorithm for hopsets for weighted directed graphs. 
The algorithm is nearly the same as the unweighted case, and so most of the analysis holds.  Our goal is to show that for a weighted graph $G$, the algorithm returns a $(n^{1/2+o(1)}, \epsilon)$-hopset of size $O(nk^{16}\log^3 n\log(nW)/\epsilon^2)$, and runs in $O(mk^{16}\log^4 n\log(nW) / \epsilon^2)$ time.  Next we will present the algorithm, and in Section \ref{section:weightedanalysis} we provide the analysis.

\subsection{Weighted Hopsets Algorithm}
Algorithm \ref{weightedhopset} shows the hopsets algorithms for weighted directed graphs.  The algorithm is the same as the unweighted algorithm with one exception.  Namely, $\textsc{WHopset}(G)$ searches all possible path weights from $-1$ to $nW$ where $W$ is the maximum weight of an edge in the graph, whereas $\textsc{Hopset}(G=(V,E))$ only searches over path weights from $n^{1/2}$ to $n$.  This difference is Line \ref{alg:line:distance}. The weighted algorithm extends the searches because the maximum shortest path distance in a weighted graph is $nW$.  In the unweighted case, the maximum shortest path was at most $n$.
$\textsc{WHopset}(G)$ searches from $-1$ to account for edges with weight zero.

\begin{algorithm*}
\caption{Hopset algorithm for weighted directed graphs. $k, \lambda$ and $L$ are parameters.}
\alglabel{weightedhopsets}
\label{weightedhopset}
\begin{algorithmic}[1]
\Function {WHopset}{$G=(V,E)$}
    \State $H \leftarrow \emptyset$
    \Loop{ $\lambda \log n$ times}
    \ForEach{$j \in [-1,\log(nW)]$} \label{alg:line:distance}
        \ForEach {$v \in V$}
            \ForEach{$i \in [0, \log_kn]$}
                \State With probability $(\lambda k^{i+1}\log n)/n$, set $\ell(v)$ to $i$, \algbreak if setting successful.
            \If{$\ell(v) \leq L$}
                \ForEach{$u \in R^{+}_{2^{j+1}}(G, v)$} add edge $(v, u)$ to $H$ with weight $\id{dist}_G(v, u)$ 
                \EndFor
                \ForEach{$u \in R^{-}_{2^{j+1}}(G, v)$} add edge $(u, v)$ to $H$ with weight $\id{dist}_G(u, v)$ 
                \EndFor
            \EndIf
            \EndFor
        \EndFor
        \State $H \leftarrow H \cup \Call{HSRecurse}{G, D=2^jk^{-c},r = 0}$
    \EndFor
    \EndLoop
    \State \Return{$H$}
\EndFunction
\end{algorithmic}
\end{algorithm*}

\subsection{Analysis} \label{section:weightedanalysis}

The goal of this section is to prove Theorem \ref{theorem:weightedhopsets}.

\begin{theorem}\label{theorem:weightedhopsets}
For any weighted directed graph $G = (V,E)$, there exists a randomized algorithm that computes a $(\beta = n^{1/2+o(1)}, \epsilon)$-hopset of size $O(nk^{16}\log^3 n\log(nW)/\epsilon^2)$ and runs in 
$O(mk^{16}\log^4 n\log(nW) / \epsilon^2)$ time with probability $1 - n^{-\lambda + 2}$.
\end{theorem}

Most of the analysis from the weighted case holds for the unweighted case.
First, we will show the difference in the runtime in Lemma \ref{lemma:weightedruntime} and then the hopbound and approximation.

\begin{lemma}\label{lemma:weightedruntime}
One execution of $\textsc{WHopset}(G=(V,E))$ with parameters $k$ and $L$, where $n=|V|$, $m=|E|$, runs in $\tilde{O}(mk^{L+1}\log(nW))$ time and returns a hopset of size $\tilde{O}(nk^{L+1}\log(nW))$ with high probability.
\end{lemma}

\begin{proof}
The running time proof follows from the proof of Theorem 
\ref{runtimetheorem}.  The only comes from performing the searches.  
Breadth-first search can no longer be used because the graph is weighted.  
Instead Dijkstra's algorithm for shortest paths  can be used which has cost $O(m + n \log n)$ \cite{Cormen:2001:IA:500824}.  
This increases the runtime from the unweighted case by a $O(\log (nW))$ factor resulting in a runtime of $O(mk^{L+1}\log^4 n \log(nW))$.  
For the same reason the size of hopset is $O(nk^{L+1}\log^3 n\log(nW))$.
\end{proof}

Next, we consider the hopbound of the weighted case.  
We again consider the path-relevant subproblems and construct the logical path-relevant subproblems.  
The only difference comes in how the logical path-relevant subproblems are constructed.
Consider a path $\hat{P}$ from $u$ to $v$, where $|w(\hat{P})| \in (k^c D / 2, k^c D]$.
If a path-relevant subproblem $(G', P, r)$ at level $r$ contains a subpath $P = \langle v_i, v_{i+1},...,v_{j}\rangle$, with $w(P) > D_r = \frac{D}{\lambda^rk^{r/2}}$ then split $P$ into $q$ disjoint subpaths $P = P_1, P_2,..., P_q$ such that $(head(P_i), tail(P_{i+1})) \in P$ for $i \in [1, q)$ and maximize each subpath $P_i$ such that $w(P_i) \leq D_r$ except for the last subpath.
Here the path is split based on weight rather than the number of hops.
Since $w(P_i) + w(head(P_i), tail(P_{i+1}) > D_r$, there are at most $\lambda^r k^{c+r/2}$ new logical nodes.
Since the rest of the number of logical nodes introduced is the same, the rest of the analysis is unaffected.

Lastly, we show the approximation of the hopsets.  For paths $P$ where $w(P) > 0$, the analysis is the same.  However for a path $P$ where $w(P) = 0$, the analysis changes.  Recall that the lightest non-zero edge weight is 1. 
The algorithm is run with $j = -1$ for this case.
When $j = -1$, we are considering the path $p$ with $w(p) < 1/2$. However, there is only $\epsilon$ error and the approximate path weight will be less than $(1+\epsilon)w(p) < 1$. Therefore, the approximate path weight is $0$ since the graph has no non-zero edge weight less than 1. 
By setting appropriate $c$, \textsc{WHopset}$(G=(V,E))$ will return a $(n^{1/2+o(1)}, \epsilon)$- hopset for $G$. 
For the final error to be $\epsilon$, set $L = 15 - 2 \log_k \epsilon$. Combining the above analysis, gives us Theorem \ref{theorem:weightedhopsets}.


\section{Parallel Algorithm} \label{section:parallel}
In this section, we show how to extend the weighted hopsets algorithm to a work-efficient, low span parallel algorithm.  First, we will explain the difficulties of the hopsets algorithm in the parallel setting and give the high-level idea of overcoming these difficulties. Then we describe the details of our parallel hopsets algorithm in Section \ref{section:parallelalg}. Finally, in Section \ref{section:parallelanalysis}, we provide an analysis of the work and span.

There are two main difficulties in making the weighted algorithm work in a parallel setting. 
First, Dijkstra's algorithm is used to perform the searches, but Dijkstra's algorithm is expensive in the parallel setting.
To resolve this problem, we use the rounding technique from Klein and Subramanian \cite{Klein:1997:RPA:266956.266959}.
Consider a path from $v_0$ to $v_{\ell}$, $\hat{P} = \langle v_0, v_1,...,v_\ell \rangle$. 
For each edge $e \in \hat{P}$, $w(e)$ is rounded up to the nearest integer multiple of $\delta w(\hat{P}) / \ell$, where $\delta$ is a small number to be set later.
Since $\hat{P}$ contains $\ell$ edges, each edge has at most $\delta w(\hat{P}) / \ell$ error.
The whole path has at most $\frac{\delta w(\hat{P})}{\ell} \cdot \ell = \delta w(\hat{P})$ error.  The error is tolerable if $\delta$ is set to be small enough.
Now consider the path with the rounded weights, but treating $\delta w(\hat{P}) / \ell$ as one unit.
Since all rounded edge weights are integer multiples of $\delta w(\hat{P}) / \ell$, the new weight of path $P$ is at most $\tilde{w}(\hat{P}) = \frac{w(\hat{P}) + \delta w(\hat{P})}{\delta w(\hat{P}) / \ell} = (1+\delta) \ell / \delta$. 
Therefore, the algorithm can use breadth first search with depth at most $O(\ell/\delta)$ to compute
$R^{+}_{\rho_vD_r}(G ,v)$ and $R^{-}_{\rho_vD_r}(G ,v)$ in a call to \textsc{HSRecurse}$(G,D = O(\ell / \delta), r)$.
The cost of the depth-first search depends only on $\ell$ instead of $w(\hat{P})$.

The second difficulty is that searching the entire path can be too expensive, even after the rounding step because a path may contain too many hops.
The key idea is to run \textsc{HSRecurse}$(G,D,r)$ with limited hops $D$.  Then add the edges produced by the \textsc{HSRecurse}$(G,D,r)$ to the graph.  Consider \textsc{HSRecurse}$(G,D,r)$ searches for at most $2\beta$ hops,  where $\beta$ is the hopbound \textsc{HSRecurse}$(G,D,r)$ achieves, and a path $\hat{P}$ with $|\hat{P}| = 4\beta$.
After the first execution of \textsc{HSRecurse}$(G,D,r)$, there will be an approximate path $P'$ for $P$ such that $|P'| \leq 2\beta$ and $w(P') \leq (1 + \epsilon) \tilde{w}(P) \leq (1+ \delta)(1+\epsilon) w(P)$.  
By repeating these steps, we can ensure that a path of any length gets approximated, and the hopbound is limited by the previous executions of \textsc{HSRecurse}$(G,D,r)$.  Moreover, for a path $P$ of any length, run \textsc{HSRecurse}$(G,D, r = 0)$ $\log(|P| / (2\beta))$ times.  
This gives a $(1+ \delta)^{\log(|P| / 2\beta)}(1+\epsilon)^{\log(|P| / 2\beta)}$ approximation, with hopbound $2\beta$.  
One more execution gives the $\beta$ hopbound.

\begin{algorithm*}
\caption{Parallel hopset algorithm for weighted directed graphs. $\delta, k, \lambda, c, L$ are parameters. }
\alglabel{hopset3}
\label{parallelalgorithm}
\begin{algorithmic}[1]
\Function {PHopset}{$G=(V,E)$}
    \State $H \leftarrow \emptyset$
    \State $\beta \leftarrow 6\lambda^{\log_kn}n^{1/2}/ \log n$
    \Loop{ $\lambda \log^2 n$ times}
        \ForEach{$i \in [-2, \log(n^2W)]$}
        \State $\hat{w} = \delta \cdot 2^{i-1} / \beta, \hat{H}' \leftarrow \emptyset$ \label{alg:line:roundstart}
        \State Construct a new graph $\hat{G} = (\hat{V} = V , \hat{E} = E)$
        \ForEach{$e \in \hat{E}$}
        \State $\tilde{w}(e) = 
            \begin{cases}
            +\infty  &\textmd{if }w(e) \geq 2^{i+1}\\
            \ceil{\frac{w(e)}{\hat{w}}} &\textmd{if }w(e) < 2^{i+1}\\
            1 &\textmd{if }w(e) = 0
            \end{cases}$ \label{alg:line:roundend}
        \EndFor
        \ForEach {$v \in \hat{V}$}
            \ForEach{$i' \in [0, \log_kn]$}
                \State With probability $(\lambda k^{i'+1}\log n)/n$, set $\ell(v)$ to $i'$, \algbreak if setting successfully.
            \EndFor
            \If{$\ell(v) \leq L$}
                \ForEach{$u \in R^{+}_{8(1 + \delta)\beta / \delta}(G, v)$} add edge $(v, u)$ to $\hat{H}'$ with weight $\id{dist}_{\hat{G}}(v, u)$ 
                \EndFor
                \ForEach{$u \in R^{-}_{8(1 + \delta)\beta / \delta}(G, v)$} add edge $(u, v)$ to $\hat{H}'$ with weight $\id{dist}_{\hat{G}}(u, v)$ 
                \EndFor
            \EndIf
        \EndFor
        \State $H \leftarrow H \cup (\hat{w} \cdot \hat{H}') \cup (\hat{w} \cdot \Call{HSRecurse}{\hat{G}, D = 4(1 + \delta) \beta /(\delta k^c ) ,r = 0} ) $ \label{alg:line:recurse}
        \EndFor
        \State $E \leftarrow E \cup H$ \label{alg:line:add-edges}
    \EndLoop
    \State \Return{$H$}
\EndFunction
\end{algorithmic}
\end{algorithm*}

\subsection{Algorithm Description} \label{section:parallelalg}
In this section, we describe the  parallel algorithm, \textsc{PHopset}$(G)$, shown in Algorithm \ref{parallelalgorithm}. 
The parallel algorithm extends the hopsets algorithm for weighted graphs in \ref{section:weighted}.
There are two main differences. 
First, the parallel algorithm will round the weights of edges.  
Second, the parallel algorithm will execute the recurse subroutine \textsc{HSRecurse}$(G',D,r)$ and then add the edges returned from the subroutine to the graph before executing the recursive subroutine again.
We will describe these two steps in more detail.

One key modification to Algorithm \ref{parallelalgorithm} is as follows.
In Lines \ref{alg:line:recurse}-\ref{alg:line:add-edges}, if the weight of an edge is less than 1, then set the weight to 0.
Also, notice that the algorithm searches from $i = -2$.  These steps are both done to account for zero weighted paths.

\paragraph{Rounding the edge weights.} The algorithm starts by rounding up the weights of edges.  This is Lines \ref{alg:line:roundstart}-\ref{alg:line:roundend} in \textsc{PHopset}$(G=(V,E))$.  Recall that the lightest non-zero edge weight is $1$, and the heaviest edge weight is $W$. $\beta$ is the hopbound of the hopset produced by the sequential algorithm \textsc{Hopset}$(G)$ in Section \ref{section:hopbound}. 

Consider a path $\hat{P} = \langle v_0, v_1,...,v_\ell \rangle$ and suppose $\ell \in (\beta, 2\beta]$ and $w(\hat{P}) \in [2^{i}, 2^{i+1})$ for integer $i$. Let $\delta$ be a small number.  Define $\hat{w} = 2^{i-1} \delta / \beta$.  Round the weight of each edge $e$ to the following integers,
\begin{equation*}
\tilde{w}(e) = 
\begin{cases}
\hat{w} & \textmd{if } w(e) = 0, \\
\ceil{\frac{w(e)}{\hat{w}}} \cdot \hat{w} 
& \textmd{if } w(e) < 2^{i+1},\\
+\infty & \textmd{if }w(e) \geq 2^{i+1}.\\
\end{cases}
\end{equation*}
By construction each edge has at most $\hat{w}$ error.  Therefore, the rounded weight of the path, $\tilde{w}(\hat{P})$ has at most $\ell \hat{w} \leq \frac{2^{i-1}\delta}{\beta} \cdot 2\beta \leq \delta d$ error.
By treating $\hat{w}$ as one unit, $\hat{P}$ is in the range of 
\begin{align*}
    \tilde{w}(\hat{P}) & \in \Bigg[ \ceil{\frac{w(\hat{P})}{\hat{w}}} \cdot \hat{w},   \ceil{\frac{(1 + \delta)w(\hat{P})}{\tilde{w}}} \cdot \hat{w}\Bigg) \\
    \subset& \Bigg[\ceil{\frac{2\beta}{\delta }} , \ceil{\frac{4(1 + \delta)\beta}{\delta}}\Bigg) \cdot \hat{w} \subset [k^c D / (2+2\delta), k^c D] \cdot \hat{w},
\end{align*}
if $k^c D = 4(1 + \delta)\beta / \delta $.
Since $\hat{w}$ is treated as one unit, breadth first search can be run to depth at most $4(1 + \delta)\beta / \delta$ to search the whole path, which is independent of $d$. In the algorithm, $\hat{w}$ is ignored in the rounding step and added back when \textsc{HSRecurse}$(G, D, r)$ returns the hopset. 

\paragraph{Adding hopset edges to the graph.} 
After the call to \textsc{HSRecurse}$(G,D,r)$, Line \ref{alg:line:add-edges} adds the edges returned by \textsc{HSRecurse}$(G,D,r)$ to the original graph $G$.
\textsc{HSRecurse}$(G,D,r)$ returns $(\beta,\epsilon)$-hopsets for any path with length at most $2\beta$ with probability at least $1/2$.  Therefore, for any path $P$ with $|P| > 2\beta$, there will be a path $P'$ approximating $P$, with length $|P'| = max(|P|/2, 2\beta)$.

\subsection{Parallel Hopbound and Hopset Size} \label{section:parallelanalysis}

\begin{lemma}
\label{parallelHopsetsBound}
Consider any graph $G' = (V,E)$ and an execution of \textsc{PHopset}$(G')$. For any $P$ where $|P|\leq 2\beta$, after the rounding code Line 6-9, suppose \textsc{HSRecurse}$(G,D,r = 0)$ returns a $(1 + \epsilon')$ approximate path $P'$ containing at most $\beta$ hops  with probability at least $1/2$. If Lines 5-16 in \textsc{PHopset}$(G')$ are repeated $j \lambda \log n$ times, then for any $u$-to-$v$ path $\hat{P}$ with $|\hat{P}| = 2^j\beta$, there will be an approximate path $\hat{P}'$ in $E$ with probability $1 - (2^j - 1)n^{-\lambda}$ such that $|\hat{P}'| \leq \beta$ and $w(\hat{P}') \leq (1+\delta)^j(1+\epsilon')^j w(P)$.
\end{lemma}

\begin{proof}
Proof by induction on $j$. When $j = 1$, then for $\hat{P}$ with $|\hat{P}| \leq 2\beta$, after $\lambda \log n$ repetitions of Lines 5-17, with all possible values of $D$, with probability $1 - \frac{1}{2^{\lambda \log n}} = 1 - n^{-\lambda}$, \textsc{HSRecurse}$(G,D,R)$ returns a $(1+\epsilon)$-approximate path for $\hat{P}$.
When the edges of $\hat{P}$ are rounded, there is at most $\delta w(\hat{P})$ error for $\hat{P}$. Therefore, the final approximation ratio is $(1+\delta)(1+\epsilon')$.

For the inductive step, we will show that for $\hat{P}$ with $|\hat{P}| \leq 2^{j+1}\beta$, the claim holds. Split $\hat{P}$ into two subpaths, $\hat{P}_1$ and $\hat{P}_2$, where each of $\hat{P}_1$ and $\hat{P}_2$ contains no more than $2^j \beta$ edges. By the inductive hypothesis, with probability $1 - (2^{j+1} - 2)n^{-\lambda}$, there exists $\hat{P}^{'}_1$ and $\hat{P}^{'}_2$ such that $\hat{P}^{'}_1$ and $\hat{P}^{'}_2$ are $(1+\delta)^j(1+\epsilon')^j$-approximations for $\hat{P}_1$ and $\hat{P}_2$, respectively. Furthermore, $|\hat{P}^{'}_1| \leq \beta$ and $|\hat{P}^{'}_2| \leq \beta$. Hence, after $\log n$ repetitions, with probability $1 - n^{-\lambda}$, there will be a $(1+\delta)(1+\epsilon')$ approximate path $\hat{P}^{'}$ with at most $H$ edges for $\langle \hat{P}^{'}_1, \hat{P}^{'}_2 \rangle$, which implies the approximate path for $\hat{P}$. By taking a union bound over the existence of $\hat{P}^{'}_1$, $\hat{P}^{'}_2$ and $\hat{P}^{'}$, the probability is $1 - (2^{j+1} - 1)n^{-\lambda}$.
\end{proof}

For $P$ with $|P| \leq 2\beta$, \textsc{HSRecurse}$(G, D, r)$ with corresponding $\hat{w}$ returns a $(\beta, \epsilon')$-hopset for $P$. 
By setting $k^c = \frac{\lambda^Lk^{(L-1)/2}}{32\log^3 n}$ and $L = 15 - 2 \log_k\epsilon'$,
the hopbound is $\beta = 6\lambda^{\log_k n}n^{1/2} / \log n$. By repeating Lines 5-16 $\lambda \log^2 n$ times, Lemma \ref{parallelHopsetsBound} can be applied to all possible paths. 
The maximum path weight will increase each round, but it will be no greater than $(1+\epsilon)^{\log n}nW \leq n^2 W$.  Thus a maximum path weight of $n^2W$ covers all possible paths. 
Finally, to get a $(\beta, \epsilon)$-hopset, set $\delta = \epsilon / (8\log n)$ and $\epsilon' = \epsilon / (8\log n)$. $L = 17 -\log_k\epsilon$ is enough if $k = \Omega(\log n)$. 
The constant $1/8$ in $\epsilon'$ will cancel out with the $\lambda^{-L}$ in the error formula. Recall that \textsc{HSRecurse}$(G,D,r = 0)$ will returns a hopset of size $O(nk^{L+1}\log^2 n)$. 
Summing up all items, the final hopset size is $O(nk^{18}\log^4 n \log(nW) /\epsilon^2)$.

\begin{corollary}
\label{parallelhopsetbound}
For any weighted directed graph $G = (V,E)$, \textsc{PHopset}$(G)$ with above parameter returns a $(\beta = n^{1/2+o(1/\log k)}, \epsilon)$-hopset of size $O(nk^{18}\log^4 n \log(nW) /\epsilon^2)$ with probability $1 - n^{-\lambda + 3}$.
\end{corollary}

\subsection{Work and Span}
Here we consider \textsc{PHopset}$(G)$ in the work-span model \cite{Cormen:2001:IA:500824}.  Recall that the work is the total number of operations that the algorithm performs while the span is the longest chain of sequential dependent operations.

\paragraph{Work.}
The work of the algorithm is dominated by the cost of the searches.  
Updating the graph, and adding the edges back to the graph can be done using parallel merge sort \cite{Cole:1988:PMS:64978.61987}.  See Fineman \cite{Fineman18} and JLS \cite{Stanford19} for details of the parallel implementation.
From the proof of Theorem \ref{runtimetheorem}, the total amount of work to compute the set of related nodes in a call of \textsc{HSRecurse}$(G,D,r)$ is $O(m k^{L+1}\log^4 n)$.
In the parallel algorithm, the $m$ term increases as more edges are added to the graph.  
When Lines 5-14 are repeated $j$ times, there are at most $O(jnk^{18}\log^2 n \log(nW) /\epsilon^2)$ edges in $H$.
The total work is,
\begin{align*}
    &O(\sum^{\lambda \log^2 n}_{j=1}(m + jnk^{18}\log^2 n \log(nW) /\epsilon^2) k^{18} \log^2 n \log(nW) / \epsilon^2) \\
    =& O(mk^{18}\log^4 n\log(nW) / \epsilon^2 + nk^{36}\log^6 n\log^2(nW) / \epsilon^4).
\end{align*}



\paragraph{Span.} The searches dominate the span.  In each call to \textsc{HSRecurse}$(G, D, r = 0)$, the maximum search distance is $4(1 + \delta)\beta/\delta$.  On each recursive call, the search distance decreases by at least 1/2.  Therefore the span in one call to \textsc{HSRecurse}$(G,D,r=0)$ is $O(\beta / \delta)$.
Since the algorithm is run $O(\log^2 n)$ times, the span is $O(\beta \log^2 n / \delta) = n^{1/2 + o(1/\log k)} \log^2 n / \epsilon$.

Summing up all these together, allows us to prove the following theorem.

\begin{theorem}\theoremlabel{Parallelruntimes}
For any weighted directed graph $G = (V,E)$, there exists a randomized parallel algorithm for weighted graph that computes a $(n^{1/2+o(1)}, \epsilon)$-hopset of size $O(n\log^{22} n \log(nW) /\epsilon^2)$ in $O(m\log^{22} n\log(nW) / \epsilon^2 + n\log^{42} n\log^2(nW) / \epsilon^4)$ work and span $n^{1/2+o(1)}/\epsilon$ with high probability.
\end{theorem}


\begin{proof}
Combining above analysis and Corollary \ref{parallelhopsetbound}, the theorem holds with $k = \Theta(\log n)$ and appropriate  $\lambda$. 
\end{proof}

\begin{theorem}
For any graph $G$ with non-negative edge weights there exists a parallel algorithm that solves approximate single-source shortest paths in $\tilde{O}(m \log(nW) / \epsilon^2 + n\log^2(nW) / \epsilon^4)$ work and span $n^{1/2+o(1)} /\epsilon$.
\end{theorem}

\begin{proof}
By \theoremref{Parallelruntimes} running \textsc{PHopset}$G$, produces a $(n^{1/2+o(1)},\epsilon)$-hopset with the desired work and span.  Then running Klein and Subramanian's hop-limited parallel algorithm for shortests paths \cite{Klein:1997:RPA:266956.266959} completes the proof.
\end{proof}

\bibliographystyle{plain}
\bibliography{bib}

\end{document}